\documentclass[twocolumn,aps,prd,amsmath,amssymb,nofootinbib]{revtex4-1}\newcommand{\cw}{\columnwidth}\newcommand{\htn}{6.7cm}\newcommand{\preprintnumber}{\hfill MIT-CTP 4573\,\,\,\,\maketitle}

\usepackage{graphicx,bm,enumerate}

\newcommand{\beq}{\begin{equation}}
\newcommand{\bea}{\begin{eqnarray}}
\newcommand{\eeq}{\end{equation}}
\newcommand{\eea}{\end{eqnarray}}
\newcommand{\mpl}{M_{Pl}}

\newcommand{\ed}{\varepsilon}

\newcommand{\nd}{n}

\newcommand{\ex}{q}
\newcommand{\F}{\Lambda}
\newcommand{\Pv}{r}
\newcommand{\ep}{\epsilon}
\newcommand{\gam}{\gamma}
\newcommand{\Mq}{B}
\newcommand{\MA}{A}
\newcommand{\Ml}{C}
\newcommand{\phim}{\rho}
\newcommand{\sig}{\sigma}

\newcommand{\N}{\mathcal{N}}
\newcommand{\Num}{N}
\newcommand{\bre}{\lambda_b}
\newcommand{\pb}{n}

\newcommand{\emo}{z}
\newcommand{\nmo}{w}
\newcommand{\Var}{\chi}

\newcommand{\kp}{k_{p}}

\newcommand{\lat}{\Delta x}
\newcommand{\lbx}{L}
\newcommand{\As}{A}
\newcommand{\td}{t_d}

\begin{document}

\title{A THEORY OF SELF-RESONANCE AFTER INFLATION\\
\vspace{0.1cm}
Part 2: Quantum Mechanics and Particle-Antiparticle Asymmetry}

\begin{abstract}
We further develop a theory of self-resonance after inflation in a large class of models involving multiple scalar fields. 
We concentrate on inflaton potentials that carry an internal symmetry, but also analyze weak breaking of this symmetry. 
This is the second part of a two part series of papers. Here in Part 2 we develop an understanding of the resonance structure from the underlying many particle quantum mechanics. We begin by a small amplitude analysis, which obtains the central resonant wave numbers, and relate it to perturbative processes. We show that the dominant resonance structure is determined by (i) the nonrelativistic scattering of many quantum particles and (ii) the application of Bose-Einstein statistics to the adiabatic and isocurvature modes, as introduced in Part 1 \cite{Part1}. Other resonance structure is understood in terms of annihilations and decays. We setup Bunch-Davies vacuum initial conditions during inflation and track the evolution of modes including Hubble expansion. In the case of a complex inflaton carrying an internal $U(1)$ symmetry, we show that when the isocurvature instability is active,  the inflaton fragments into separate regions of $\phi$-particles and anti-$\phi$-particles. We then introduce a weak breaking of the $U(1)$ symmetry; this can lead to baryogenesis, as shown by some of us recently \cite{Hertzberg:2013jba,Hertzberg:2013mba}. Then using our results, we compute corrections to the particle-antiparticle asymmetry from this preheating era.
\end{abstract}

\author{
Mark P.~Hertzberg$^*$, 
Johanna Karouby$^\dagger$, 
William G.~Spitzer,
Juana C.~Becerra, 
Lanqing Li
}
\affiliation{Center for Theoretical Physics and Dept.~of Physics,\\ 
Massachusetts Institute of Technology, Cambridge, MA 02139, USA}

\let\thefootnote\relax\footnotetext{$^*$Electronic address: {\tt mphertz@mit.edu}\\$^\dagger$Electronic address: {\tt karoubyj@mit.edu}}

\date{\today}

\preprintnumber

\maketitle

\tableofcontents

\section{Introduction} \label{Introduction} 

Inflationary cosmology provides an account of several otherwise puzzling features of the universe, namely the large scale homogeneity, isotropy, and flatness \cite{Guth:1980zm,Linde:1981mu,Albrecht:1982wi}. Recent observations are in good agreement with the basic predictions of inflation, including a nearly scale-invariant spectrum of primordial fluctuations, Gaussianity, etc. Recent tantalizing evidence of primordial B-modes \cite{BICEP} would provide information about the inflationary energy scale. Altogether, although the full details of inflation are not known, significant progress is being made, 
both observationally \cite{WMAP,Planck,Dodelson:2009kq} and theoretically \cite{Lyth:1998xn,Kachru:2003sx,Dimopoulos:2005ac,Linde:2007fr,Hertzberg:2011rc,Hertzberg:2014aha,Baumann:2014nda,Hertzberg:2014sza,Kaiser:2013sna}.

However, the post-inflationary era is much more uncertain. This era is essential for understanding the transition from the inflaton into other fields, including the Standard Model degrees of freedom. In this post-inflationary era, tremendous power can be generated on small scales from particle interactions. Indeed various forms of resonance can take place, including self-resonance, as quantum perturbations of the inflaton are ``pumped" by the homogeneous background.

Many interesting works, often investigating the coupling to other fields, has appeared in the literature \cite{Kofman:1994rk,Kofman:1997yn,Greene:1997fu,Battefeld:2008bu,Greene:1997ge,Bassett:1999ta,Dufaux:2006ee,Greene:1998nh,Greene:2000ew,Peloso:2000hy,Davis:2000zp,Braden:2010wd,Deskins:2013dwa,Barnaby:2011qe,GarciaBellido:2008ab,Figueroa:2009jw,Ashoorioon:2013oha,Amin:2010dc,Gleiser:2011xj,Amin:2011hj,Hertzberg:2010yz,Karouby:2011xs}. 
This includes Refs.~\cite{Kofman:1994rk,Kofman:1997yn}, which emphasized a coupling of the inflaton $\phi$ to a daughter field $\chi$, with interactions such as $\sim g^2\phi^2\chi^2$ or $\sim g\,\phi\,\chi^2$. Under certain circumstances, this can cause a dramatic growth in $\chi$, that goes beyond standard perturbation theory. Other interactions include coupling to gauge fields \cite{Davis:2000zp,Deskins:2013dwa}, coupling to fermionic fields \cite{Greene:1998nh,Greene:2000ew}, and the metric itself \cite{Bassett:1999ta}.
Self-resonance, where the inflaton pumps its own fluctuations, can occur for potentials with nonlinearities, including the quartic term $\sim\lambda\,\phi^4$, as discussed in \cite{Greene:1997fu}. In some parameter regimes (namely negative $\lambda$), this can produce an abundance of coherent structures, such as oscillons; see \cite{Amin:2010dc,Gleiser:2011xj,Amin:2011hj}.
In this work, we focus on the important issue of self-resonance of the inflaton, and assume couplings to other fields are small. 
We will understand the structure of the self-resonance from the point of view of many particle quantum mechanics and apply the results to a model of baryogenesis; which appears to go beyond the existing literature.

During this phase, a detailed understanding of the relationship between the classical field approximation and the quantum behavior of many particles is important. But perhaps the most important feature of the early universe that remains uncertain is the generation of all the matter in the universe. This is thought to arise from the decay of the inflaton. If subsequent interactions are sufficiently symmetric between particles and antiparticles, then no net baryon number will be left over. Hence it is essential to formulate models of the generation of asymmetry between particles and antiparticles. In this paper we address these quantum and particle-antiparticle asymmetry issues.

This is the second in a two part series of papers. In the first paper \cite{Part1} we introduced a large class of interesting models. Namely, models with an arbitrary number of scalar fields, organized by an internal $\mathcal{O}(\N)$ symmetry. Since couplings to other fields can be small, (as is often assumed for the flatness of the inflationary potential to be technically natural), it can sometimes be the case that self-resonance after inflation is most important. With multiple fields, we showed in Part 1 \cite{Part1} that the field decomposes into adiabatic and isocurvature modes. We showed that the spectrum is gapless, as required by the Goldstone theorem,
and derived the growth rates (``Floquet exponents") from the appropriate time averaged pressure and densities. We saw that the resonance structure could be particularly efficient at relatively long wavelengths, as it is dominated by the first instability band. We found that for positive self couplings, the adiabatic mode is stable, while the isocurvature modes are unstable. While for negative self couplings, the adiabatic mode is unstable, while the isocurvature modes are stable. This was derived from the background pressure associated with the adiabatic mode, and an auxiliary pressure associated with the isocurvature modes.

In this second paper, we introduce quantum mechanics in two important respects to extend the classical field theory analysis of Part 1 \cite{Part1}. Firstly, we understand the stability structure from the point of view of many particle quantum mechanics. The behavior of adiabatic and isocurvature modes at long wavelengths can be understood in terms of nonrelativistic quantum mechanics. We describe the (sometimes subdominant) higher instability bands perturbatively, using Feynman diagrams involving annihilation and decays of the parent inflaton into relativistic daughter particles. As an important stepping stone to this analysis, we first perform a (classical) small amplitude analysis, which connects to the Feynman diagrams directly.
Secondly, we quantize the inflationary fields. We put the inflaton in its Bunch-Davies vacuum initial conditions. We then track the modes under Hubble expansion. We show how the resonant modes can grow approximately exponentially in the slow redshifting regime. We compute the final power spectra of adiabatic and isocurvature modes.

These spectra set the probability distributions for the fields. We draw from these probability distributions. For the case of two fields, with a $U(1)$ symmetry, we find that when the isocurvature instability is active, the inflaton fragments into separate regions of particles and antiparticles. In this way, the symmetry between particles and antiparticles is {\em spontaneously} broken.

In the case of a complex inflaton, we go further and introduce an {\em explicit} breaking of the $U(1)$ symmetry. In some models, the breaking can lead to an over abundance of inflaton particles over antiparticles (or vice versa). This may further lead to the cosmological baryon asymmetry if the inflaton can decay into quarks appropriately; as showed by some of us recently in Refs.~\cite{Hertzberg:2013jba,Hertzberg:2013mba}. Here we include the leading corrections from self-resonance. We show how to use the symmetric theory to obtain these leading corrections. The asymmetry is found to be proportional to an integral over the difference in power spectra between the adiabatic and isocurvature modes. So while each of these is individually UV divergent, the difference leads to a finite contribution.

The outline of this paper is as follows:
In Section \ref{Model} we present the class of models under investigation and recap numerical results for dimension 4 potentials.
In Section \ref{SmallAmplitude} we derive analytical results for small inflaton amplitudes, for both the first and second instability bands.
In Section \ref{Quantum} we discuss the connection of our results to the quantum mechanics of many particles.
In Section \ref{Hubble} we include Hubble expansion in the analysis.
In Section \ref{PositionSpace} we quantize the fields and sample the ground state wavefunctionals to present the fields in position space.
In Section \ref{Baryogenesis} we apply our results to inflationary baryogenesis models.
Finally, in Section \ref{Conclusions} we discuss our findings and conclude.

\section{Symmetric and Asymmetric Theories} \label{Model} 

Many high energy particle physics models involving one or more scalar fields coupled to gravity. Since scalar fields can, under appropriate conditions, lead to an effective equation of state $w\approx-1$, then they can lead to a period of inflation. As in Part 1 \cite{Part1}, we consider $\N$ scalar fields and organize them into a vector
\beq
\vec{\phi}=\{\phi_1,\ldots,\phi_\N\}
\eeq
Later we will specialize to the case of two scalar fields. In that case it is particularly convenient to organize them into a complex scalar as follows
\beq
\phi = {\phi_1+i\phi_2\over\sqrt{2}}
\eeq

We now discuss the structure of the dynamics. Since we will emphasize quantum effects in this paper, it is appropriate to recall that the inflationary action is only an effective field theory, since gravitation is non-renormalizable in 4 dimensions. However, in a weakly coupled model, corrections from the leading two-derivative action are typically small (though exceptions are possible). So here we assume, for simplicity, that all higher order derivative corrections to the Einstein-Hilbert action are small. Furthermore, we specialize to canonical kinetic energy in the Einstein frame. Since inflaton couplings are usually small in order to achieve small fluctuations, this is technically natural.
So we take the action for $\N$ scalar fields to be (signature $-+++$, units $\hbar=c=1$)
\beq
S=\int d^4x\sqrt{-g}\left[{\mpl^2\over2}\mathcal{R}-{1\over2}\delta_{ij}\,\partial_\mu\phi^i\partial^\mu\phi^j-V(\vec\phi)\right]\,\,
\eeq
where $\mpl\equiv1\sqrt{8\pi G_N}$ is the reduced Planck mass. In the Appendix of Part 1 \cite{Part1} we developed some results for more general potentials, including higher derivative corrections and non-trivial metrics $G_{ij}$ on field space. But these generalizations are not important for our analysis here.

Ignoring higher order corrections, the residual freedom is in the choice of the potential $V(\vec\phi)$. In Part 1 \cite{Part1} we exclusively studied symmetric potentials that carry the internal rotational symmetry
\beq
\phi^i\to R^i_j\,\phi^j
\eeq 
where $R$ is a rotation matrix acting on field space. Formally this implied an $O(\N)$ symmetry and the potential may be written as $V(\vec\phi)=V(|\vec\phi|)$. Here we often focus on these symmetric potentials, but we also allow for a breaking of the symmetry. So we decompose the potential as
\beq
V(\vec\phi) = V_s(|\vec\phi|)+V_b(\vec\phi)
\eeq
The term $V_s$ is a symmetric potential that carries the internal rotational symmetry, while $V_b$ does not. For most of this paper we ignore the breaking term and utilize the symmetry to simplify the analysis. We then make use of the breaking term in Section \ref{Baryogenesis} to provide an asymmetry between particles and antiparticles. This breaking of a symmetry means that particle number is not exactly conserved, potentially leading to the matter-antimatter asymmetry.

\subsection{Classical Evolution (Preliminary)}\label{Background}

We begin by discussing the evolution of the classical background. The metric is established by inflation to be the standard flat FRW metric
\beq
ds^2 = -dt^2 +a(t)^2 d{\bf x}^2
\eeq
where $a(t)$ is the scale factor. 

The evolution of the classical field is, in general, complicated. When the symmetry breaking term $V_b$ is present, the field tends to get kicked around in field space. If the breaking is small, then it is roughy a kind of elliptic behavior. This will play a role later in our Section \ref{Baryogenesis} on baryogenesis.

On the other hand, if the breaking term is negligible, then the motion simplifies considerably. Inflation tends to erase angular momentum, even in field space. This leads to the multi-field inflaton moving radially in fields space.  As in Part 1 \cite{Part1}, the purely radial motion for the background shall be denoted by the field $\phi_0(t)$. 
From varying the action, we obtain the standard equation of motion for a homogeneous scalar field
\beq
\ddot\phi_0+3H\dot\phi_0+V'(\phi_0) = 0
\eeq
where $H=\dot a/a$ is the Hubble parameter. During slow-roll inflation, the second and third terms here dominate. After inflation, as is the focus of this work, the first and third terms dominate and the second ``friction" term is sub-dominant.

In Section \ref{Hubble} we will properly track the behavior for $\phi_0$, where we self consistently solve for the Hubble parameter $H$ from the Friedmann equation
\beq
H^2 = {1\over 3\mpl^2}\left({1\over2}\dot\phi_0^2+V_s(\phi_0)\right)
\eeq
This leads to a redshifting of the background (classical) fields that will influence the self-resonance of the perturbations in an important fashion.

\subsection{Quantal Evolution (Preliminary)} \label{Linearized}

Due to quantum mechanics the field cannot have a well defined value, so there are necessarily quantum fluctuations.
Focussing then on the symmetric case, we can decompose these fluctuations into those that are parallel to the radial motion of the background $\delta\phi_\parallel$, and those that are orthogonal to the background $\delta\phi_\perp$. Later in Section \ref{Quantum} we will be precise about the quantization of these fluctuations. But for now it suffices to treat them as any form of fluctuation, either classical or quantum, even though its origin is necessarily quantum.
We expand the field around the background as 
\beq
\vec\phi({\bf x},t) = \vec\phi_0(t)+\delta\vec\phi({\bf x},t)
\eeq
where
\bea
\delta\vec\phi({\bf x},t) = \{\delta\phi_{\perp 1}({\bf x},t),\ldots,\delta\phi_{\perp \N-1}({\bf x},t),\delta\phi_\parallel({\bf x},t)\}\,\,\,\,\,\,
\eea
and we have put the background motion in the $\N$th direction, without loss of generality, in the symmetric theory. Later in Section \ref{Baryogenesis} we will allow for a general direction for the asymmetry theory.

Expanding the perturbations (classical or quantal) to first order we have
\bea
&&\,\ddot{\delta\phi_\parallel}+3H\dot{\delta\phi_\parallel}+\left({k^2\over a^2}+V''(\phi_0)\right)\delta\phi_\parallel=\mathcal{G}\label{PertEqn1}\\
&&\ddot{\,\delta\phi_{\perp i}}+3H\!\dot{\,\delta\phi_{\perp i}}+\left({k^2\over a^2}+{V'(\phi_0)\over\phi_0}\right)\delta\phi_{\perp i}=0\label{PertEqn2}
\eea
where we have Fourier transformed to $k$-space. For the orthogonal components, we have included an ``$i$" index, where $i$ runs over $i=1,\ldots,\N-1$; each equation carries the same structure due to the symmetry. 

We note that at linear order one can include linearized corrections to the metric, such as the Newtonian potential. It is possible to make a gauge choice in which the fluctuating dynamical degrees of freedom are solely described by the scalar field fluctuations. In the gauge in which there are spatially flat hypersurfaces, the function $\mathcal{G}$ is
\beq
\mathcal{G}={1\over a^3\mpl^2}{d\over dt}\!\left(a^3\dot\phi_0^2\over H\right)\delta\phi_\parallel
\label{PertEqn1cor}
\eeq
One can show that on sub-Hubble scales, such corrections are small, in particular they are suppressed relative to the terms included in eqs.~(\ref{PertEqn1},\,\ref{PertEqn2}) by $\sim a^2H^2/k^2$. On the other hand, for order Hubble or super-Hubble scales, such corrections can be important. For simplicity, we ignore such corrections in this work, and we suspect this will not change our central conclusions. In fact we will see that our primary effects occur on lengths scales that are not parametrically larger than the Hubble length. So this simplification is reasonable.

\begin{figure*}[t]
  \includegraphics[width=\textwidth]{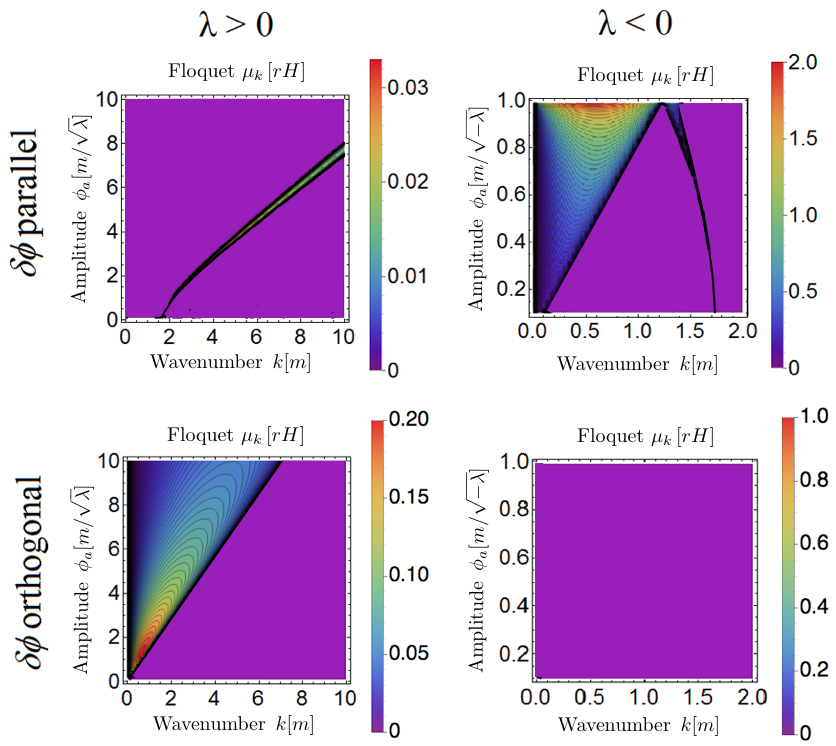}
  \caption{Contour plot of the real part of Floquet exponent $\mu_k$ for dimension 4 potentials as a function of wavenumber $k$ and background amplitude $\phi_a$ with $m^2>0$. Left panel is $\lambda>0$ and right panel is $\lambda<0$. Upper panel is $\delta\phi_\parallel$ and lower panel is $\delta\phi_{\perp}$. We have plotted $\mu_k$ in units of $\Pv H$ where $\Pv\equiv \sqrt{|\lambda|}\,\mpl/m$, $k$ in units of $m$, and $\phi_a$ in units of  $m/\sqrt{|\lambda|}$. (This is taken from Part 1 \cite{Part1}.)}
  \label{4PanelView}\end{figure*}

\subsection{Floquet Results for Dimension 4 Potentials} \label{Quartic}

As we describe explicitly in Section \ref{QuantizePert} where we quantize the perturbations, the mode functions satisfy these classical equations of motion. In this case, it is useful to develop a numerical recipe to solve these equations. Here we provide a brief recap of the central numerical results found in Part 1 \cite{Part1}.

Importantly, we need a form for the potential $V_s$. We consider the regime well after inflation, where the potential should be well approximated by its leading order operators. Since the potential is assumed to carry an internal rotational symmetry, we can expand it as
\beq
V_s(\vec\phi) = V_0 + {1\over 2}m^2|\vec\phi|^2 +{\lambda\over 4}|\vec\phi|^4+\ldots
\eeq
For sufficiently small field amplitudes, these leading dimension 4 terms will dominate the dynamics. Such a regime will normally arise after a sufficient amount of redshifting has occurred. A counter example would be if some of the above coefficients happen to vanish; we will consider this possibility in Section \ref{SmallAmplitude}. For large amplitudes, higher order corrections to the potential may be important (we mention a toy example example in Section \ref{BaryonNumerical}).

In Part 1 \cite{Part1} we described the recipe to obtain the Floquet exponents $\mu_k$, which govern any possible exponential growth in the modes. This is rigorously defined when the background is oscillating periodically. This is a good approximation in the limit in which the oscillation time scale is short compared to the Hubble time. We will return to these details later in Section \ref{Hubble}. For now we truncate the potential to purely dimension $\leq 4$ terms and numerically solve for the corresponding Floquet exponent using the method of Part 1 \cite{Part1}.

In Fig.~\ref{4PanelView} we recap the results for the Floquet exponent from Part 1 \cite{Part1}. We have plotted the dimensionless quantity $\mu_k(\Pv H)$. Here $\Pv$ is the dimensionless parameter 
\beq
\Pv\equiv {\sqrt{|\lambda|}\mpl\over m}
\eeq
It is found to control the amount of resonance in the problem. As we show in Section \ref{Hubble}, for $\Pv\gg 1$, $\mu_k/H$ can be large and there is significant resonance, else there is rather insignificant resonance. 
The variable $\mu_k/(\Pv H)$ is convenient here as it scales out the physical parameters.
In the left panel we have taken the coupling $\lambda>0$ and in the right panel we have taken the coupling $\lambda<0$. In the upper panel we study the parallel, or ``adiabatic", perturbations. In the lower panel we study the orthogonal, or ``isocurvature", perturbations.
We see clearly that for $\lambda>0$ there is a large instability for the isocurvature mode, while for $\lambda<0$ there is a large instability for the adiabatic mode. There is also a band originating from $k=\sqrt{3}\,m$ at small amplitudes for the adiabatic mode only for either sign of $\lambda$.

\begin{figure}[t]
  \includegraphics[width=\cw]{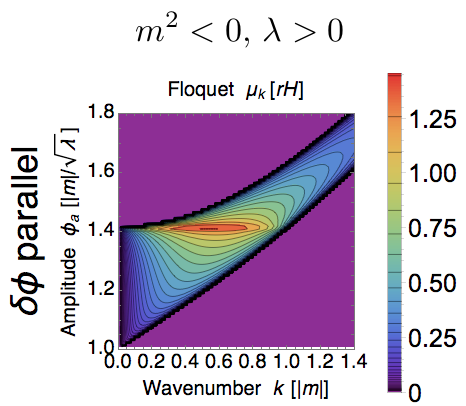}\\\vspace{0.2cm}
  \includegraphics[width=\cw]{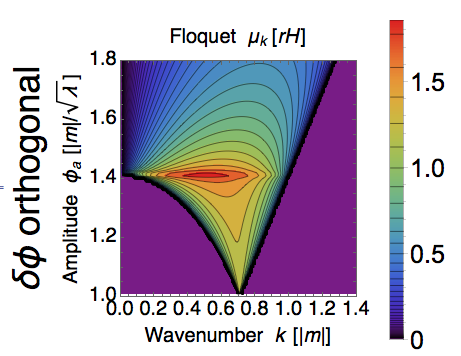}
  \caption{Contour plot of the real part of Floquet exponent $\mu_k$ for dimension 4 potentials as a function of wavenumber $k$ and background amplitude $\phi_a$ with $m^2<0$ and $\lambda>0$. Upper panel is for $\delta\phi_\parallel$ and lower panel is for $\delta\phi_{\perp}$. 
 We have plotted $\mu_k$ in units of $\Pv H$ where $\Pv\equiv \sqrt{\lambda}\,\mpl/|m|$, $k$ in units of $|m|$, and $\phi_a$ in units of  $|m|/\sqrt{\lambda}$. (This is taken from Part 1 \cite{Part1}.)} 
\label{NegativeMass}\end{figure}

Furthermore, in Fig.~\ref{NegativeMass} we allow for a Higgs type potential with $m^2<0$ and $\lambda>0$. In this case we have focussed on the just the first instability band. We see that it begins at $k=0$ for the parallel perturbations, and at $k=|m|/\sqrt{2}$ for the isocurvature perturbations. 

We will explain the above observations in this paper from the underlying quantum mechanics on many particles. As a step in this direction, we being with the analytical treatment of perturbation theory at small amplitude.

\section{Small Amplitude: Analytical Results} \label{SmallAmplitude}

In the previous section we presented numerical results on the behavior of all the perturbations, and in Part 1 \cite{Part1} we presented analytical results a long wavelengths using a pressure analysis. It is also important to have analytical results at both long and short wavelengths. This can be achieved if we focus on small amplitudes of the inflaton field; this will eventually arise after sufficient redshifting. 

At small amplitudes, we can perform a weakly coupled expansion around the almost free theory. To do so, we assume there exists a mass term which dominates the oscillatory behavior of the background giving rise to almost harmonic motion. Plus we add interactions that are sub-dominant leading to anharmonic behavior, and can possibly drive resonance in perturbations. We write the potential as an expansion as
\beq
V_s(\vec\phi) = V_0 + {1\over2}m^2|\vec\phi|^2+{\lambda\over2\,\ex}|\vec\phi|^{2\ex}+\ldots
\eeq
where we assume $\ex$ is an integer $\ex\ge2$ which governs the {\em leading} interaction term around small field values. 
Of most interest is the case $\ex=2$, giving rise to a standard dimension 4 interaction term. On general effective field theory grounds, we should expect this quartic term to exist. It is conceivable that the quartic term vanishes and that the leading interaction term begins at $\ex=3$ or higher. For now we simply allow for a general integer power, and specialize to the quartic case $\ex=2$ when necessary.

\subsection{Background Evolution}

We begin by studying the positive mass squared case ($m^2>0$). Later we shall study the negative mass squared (tachyonic) case in Section \ref{NegMass}. So for now, we set the vacuum energy $V_0 = 0$ and allow the interaction's coupling $\lambda$ to be either positive or negative.

As before, the background evolution $\vec\phi_0$ is taken to be radial. The equation of motion is
\beq
\ddot\phi_0 + m^2\phi_0+\lambda\,\phi_0^{2\ex-1} = 0
\eeq
We would like to solve this in a small amplitude expansion. So lets expand the background $\phi_0$ as
\beq
\phi_0 = \ep \, \phi_1 + \ep^{2q-1}\,\phi_{2\ex-1}+\ldots
\label{expansion}\eeq
where $\ep$ is a small dimensionless constant that sets the power counting. The functions $\phi_1,\,\phi_{2\ex-1},\ldots$ are functions of time to solve for.

Naively, we would like to substitute this expansion into the equation of motion directly and match powers of $\ep$. However this would lead to secular behavior as the driving terms would carry the same frequency as the natural frequency defined by the harmonic terms. To avoid this problem, we need to identify a shifted frequency. We do this by introducing a new time variable as follows
\beq
\tau \equiv t\sqrt{1\pm\ep^{2\ex-2}}
\eeq
where the upper ``+" sign is for $\lambda>0$, as the interaction will raise the fundamental oscillation frequency, and the lower ``-" sign is for $\lambda<0$, as the interaction will lower the fundamental oscillation frequency. This also allows us to define the value of $\ep$ uniquely, as we see shortly. So with respect to $\tau$ we have the equation of motion
\beq
{d^2\phi_0\over d\tau^2}\pm\ep^{2\ex-2}{d^2\phi_0\over d\tau^2} + m^2\phi_0+\lambda\,\phi_0^{2q-1} = 0
\eeq

Now using the expansion (\ref{expansion}) and matching powers at $\mathcal{O}(\ep)$ gives
\beq
{d^2\phi_1\over d\tau^2}+ m^2\phi_1 = 0
\eeq
We write the solution as
\beq
\phi_1 = \phi_{a1}\cos(m\,\tau)
\label{phi1}\eeq
where we have dropped an overall phase.
Here $\phi_{a1}$ is a type of amplitude. At this leading order, the full amplitude for $\phi_0$ is related to this by 
\beq
\phi_a = \ep\,\phi_{a1}
\label{phiaeps1}\eeq
Then using the expansion (\ref{expansion}) and the solution for $\phi_1$ from eq.~(\ref{phi1}) and matching powers at  $\mathcal{O}(\ep^{2\ex-1})$ gives
\bea
&&{d^2\phi_{2\ex-1}\over d\tau^2} + m^2\phi_{2\ex-1} \nonumber\\
&&= \pm m^2\phi_{a1}\cos(m\,\tau)-\lambda\,\phi_{a1}^{2\ex-1}\cos^{2\ex-1}(m\,\tau)\,\,\,\,
\label{2qm1}\eea
The right hand side acts as a {\em driving} term. We need to remove the piece that is proportional to $\cos(m\,\tau)$ as it would otherwise drive a resonance leading to secular behavior. For any integer power $\ex$, the final cosine has a leading harmonic given by
\beq
\cos^{2\ex-1}(m\,\tau) = {C(2\ex-1,\ex)\over 2^{2\ex-2}} \cos(m\,\tau)+\mbox{h.h} 
\eeq
where $C$ is the binomial coefficient and ``h.h" represents higher harmonics. 
We substitute this into (\ref{2qm1}) and demand that the coefficient of $\cos(m\,\tau)$ vanishes. This leads to a unique value for the amplitude $\phi_{a1}$, which we find to be
\beq
\phi_{a1} = 2\left[m^2\over C(2\ex-1,\ex)|\lambda|\right]^{1\over2\ex-2}
\label{phia1}\eeq
This finalizes the background solution $\phi_0$ at leading order as $\phi_0 = \ep\,\phi_1$, with $\phi_1$ given by eqs.(\ref{phi1},\ref{phia1}) and $\ep$ parameterizing the amplitude.

\subsection{First Instability Band} \label{First}

We now examine the behavior of perturbations about this background. For the sake of brevity, here we will present a unified treatment of the adiabatic and isocurvature modes, rather than separate analyses. 

By switching to the new time variable $\tau$, the Hill's equation becomes
\beq
{d^2\over d\tau^2}\delta\phi+h(\tau)\delta\phi = 0
\eeq
where $h(\tau)$ is the periodic pump with respect to the $\tau$ variable. It is given by
\beq
h(\tau) =  {k^2 + m^2+\gam\,\lambda\,\phi_0^{2\ex-2}(\tau)\over 1\pm\ep^{2\ex-2}}
\eeq
where we have divided throughout by the factor $1\pm\ep^{2\ex-2}$ to make the second derivative term in Hill's equation canonical.
In $h$ we have introduced the factor $\gam$ which distinguishes the two classes of modes as
\beq
\gam = \Big{\{}
\begin{array}{c}
2\ex-1\,\,\,\,\mbox{for}\,\,\,\delta\phi_\parallel\\
\,1\,\,\,\,\,\,\,\,\,\,\,\,\,\,\,\,\mbox{for}\,\,\,\delta\phi_\perp
\end{array}
\label{gamdef}\eeq
Since the driving term is given by $\lambda\,\phi_0^{2\ex-2}$, we would like to expand this in terms of harmonics using our leading order $\ep$ result of the previous subsection. We find a constant term and a piece promotional to $\cos(2\,m\,\tau)$ as follows
\bea
\lambda\,\phi_0^{2\ex-2} = \pm {m^2\ep^{2\ex-2}\over2\ex-1}\Big{[}\ex
	+(2\ex-2)\cos(2\,m\,\tau)+\mbox{h.h}\Big{]}\,\,\,\,\,\,\,\,\,\,\,\,\,
\eea
where again ``h.h" represents higher harmonics. Substitution into $h(\tau)$ and working to leading non-zero order $\sim\ep^{2\ex-2}$ re-organizes $h$ into the form of the so called Mathieu equation
\beq
h(\tau) = \MA+2\,\Mq\cos(2\,m\,\tau)
\eeq
where for now we drop higher harmonics; these will only be important for the second instability band that we discuss in the next subsection.
We find that the Mathieu $\Mq$ and $\MA$ coefficients are
\bea
\Mq &=& \pm\,\gam\, m^2\ep^{2\ex-2}{\ex-1\over2\ex-1} \\
\MA &=& k^2+m^2+\bar\gam\,\Mq/\gam
\eea
where 
\beq
\bar\gam = \Big{\{}
\begin{array}{c}
2\ex-1\,\,\,\,\mbox{for}\,\,\,\delta\phi_\parallel\\
-1\,\,\,\,\,\,\,\,\,\,\,\,\,\,\mbox{for}\,\,\,\delta\phi_\perp
\end{array}
\eeq
We note that in the first instability band $k\sim \ep^{\ex-1}$ so we have not included powers of $\ep$ that multiple $k^2$ when expanding out the denominator that appears in $h$.

Now the Mathieu equation can be solved by performing a harmonic expansion as follows
\beq
\delta\phi(\tau) = \sum_\omega e^{i\omega\tau}\delta\phi_\omega(\tau)
\label{harmonics}\eeq
where the frequencies are summed over integer multiplies of the mass $m$; the fundamental frequency, with $-\infty<\omega<+\infty$. 
Here we assume the $\delta\phi_\omega$ are slowly varying in time. Substitution into the Mathieu equation and matching harmonics, gives the following coupled system of ODEs
\beq
2i\omega{d\over d\tau}\delta\phi_\omega
+(\MA-\omega^2)\delta\phi_\omega+\Mq(\delta\phi_{\omega-2m}+\delta\phi_{\omega+2m}) = 0
\eeq
where we have dropped the second order derivative ${d^2\over d\tau^2}\delta\phi_\omega$ since $\delta\phi_\omega$ is slowly varying.
Notice that odd harmonics are only coupled to odd harmonics, and even harmonics are only coupled to even harmonics.

The first instability band comes from studying the fundamental frequencies $\omega=+m$ and $\omega=-m$. To leading order, these evolve independently of the higher harmonics, allowing us to truncate this system to just these frequencies. This leads to the following pair of ODEs
\beq
{d\over d\tau}\left(\!\begin{array}{c}
\delta\phi_{+m}\\
\delta\phi_{-m}
\end{array}\!\right)
={i\over2m}\left(\!\begin{array}{cc}
\MA-m^2 & \Mq\\
-\Mq & m^2-\MA
\end{array}\!\right)
\left(\!\begin{array}{c}
\delta\phi_{+m}\\
\delta\phi_{-m}
\end{array}\!\right)
\eeq
The eigenvalues of this matrix gives the Floquet exponents to leading order for small amplitudes
\beq
\mu_k = {1\over2m}\sqrt{\Mq^2-(\MA-m^2)^2}
\eeq
(we should take both signs of the square root to get both Floquet exponents).
Substitution of the above values for $\Mq$ and $\MA$ into this and eliminating $\ep$ in favor of the physical amplitude $\phi_a$ using (\ref{phiaeps1},\,\ref{phia1}), we obtain the following result for the Floquet exponent
\bea
\mu_k = {k\over2\,m}\sqrt{-\alpha\,\bar\gam\,\lambda\,\phi_a^{2\ex-2}-k^2}\,\,\,\,\,\,\,\,\,\,\,\,\,\,
\label{muSA}\eea
where
\beq
\alpha \equiv {C(2\ex-1,\ex)(\ex-1)\over 2^{2\ex-3}(2\ex-1)}
\label{alpha}\eeq
(for $\ex=2$; $-\alpha\,\bar\gam = -3/2$ for $\delta\phi_\parallel$, $-\alpha\,\bar\gam = +1/2$ for $\delta\phi_\perp$).
Since the first term inside the square root is proportional to $-\bar\gam\,\lambda$, with all other factors positive, we see that the existence of an instability band is determined by the sign of $-\bar\gam\,\lambda$. So this proves that if $\lambda>0$ there is an instability for the isocurvature mode ($\bar\gam=-1<0$) and if $\lambda<0$ there is an instability for the adiabatic mode ($\bar\gam=2\ex-1>0$). So again we find an entire class of potentials whose stability is complementary between adiabatic and isocurvature modes.

For the cases in which there is an instability band, the right hand edge of the band has the shape 
\beq
k_{r,edge}= \sqrt{-\alpha\,\bar\gam\,\lambda}\,\,\phi_a^{\ex-1}
\eeq
For the important case $\ex=2$, this gives a linear relationship between $k_{r,edge}$ and amplitude $\phi_a$. The left hand edge of the instability band is at
\beq
k_{l,edge}= 0
\eeq
which connects to the long wavelength analysis of Part 1 \cite{Part1}. For a fixed amplitude $\phi_a$, the Floquet exponent is maximized for $k_{max} = k_{r,edge}/\sqrt{2}$. The corresponding maximum Floquet exponent is
\beq
\mu_{max} = {-\alpha\,\bar\gam\,\lambda\,\phi_a^{2\ex-2}\over 4\,m}
\eeq
(so $\mu_{max}\propto \phi_a^2$ for $\ex=2$).

Altogether this explains the width and shape of the first instability bands seen earlier in Fig.~\ref{4PanelView}. Furthermore, when $\ex=2$, we see that $\mu_k$ for the adiabatic mode can be related to $\mu_k$ for the isocurvature mode by the replacement
\beq
\lambda\to-{\lambda\over3}
\eeq
This is in agreement with the result we proved in Part 1 \cite{Part1}, where we derived an auxiliary potential and Taylor expanded for small amplitudes.

Out of interest, let us take the small $k$ limit of this result. This leaves a result for $\mu_k$ that is linear in $k$, as we proved it should be in Part 1 \cite{Part1} where we did a general long wavelength analysis, and the Goldstone theorem ensured a gapless spectrum. If we are both at small wavenumber and small amplitude (the lower left corner of the stability charts) we obtain from eqs.~(\ref{muSA},\,\ref{alpha}) the following relationship among the speeds
\beq
c_I^2 = -{c_S^2\over 2\ex-1}
\eeq
whose sign governs the stability structure. Indeed for the case $\ex=2$, we obtain $c_I^2 = - c_S^2/3$.

\subsection{Second Instability Band} \label{Second}

In the previous subsection we discussed the $\ep$ expansion that governs the behavior at small amplitudes, and applied it to the first instability band. In principle one can go further and study any band to any desired order in perturbation theory. Here we mention the key leading order results of the second instability band. We continue to study $m^2>0$, but now we specialize to $\ex=2$, i.e., dimension 4 potentials.

In the previous subsection we found the solution to $\mathcal{O}(\ep)$ in eqs.~(\ref{phi1},\,\ref{phia1}) (by taking $\ex\to2$). For the next band, we need the background solution to $\mathcal{O}(\ep^3)$. Without going through the full details, we find that the result to this order is
\beq
\phi_0 = (\ep\,\phi_{a1}+\ep^3\phi_{a3})\cos(m\,\tau)+\ep^3{\lambda\,\phi_{a1}^3\over 32\, m^2}\cos(3\,m\,\tau)
\eeq
where
\beq
\phi_{a1}={2m\over\sqrt{3|\lambda|}},\,\,\,\,\,\,\phi_{a3} = \mp{m\over 24\sqrt{3|\lambda|}}
\eeq
Substitution into the perturbation equations leads to a Hill's function $h$ that is more complicated than the standard Mathieu equation, namely
\beq
h(\tau) = \MA +2\,\Mq\cos(2\,m\,\tau)+2\,\Ml\cos(4\,m\,\tau)
\eeq
where we have gone to the required number of harmonics. The coefficients are required to $\mathcal{O}(\ep^4)$, and we find them to be
\bea
\Mq &=&{\gam\,m^2\over72}\left(\pm24\ep^2-23\ep^4\right)\\
\Ml &=&{\gam\,m^2\over36}\ep^4\\
\MA &=& m^2\!\left(1\pm\left({2\gam\over3}-1\right)\ep^2+\left(1-{25\gam\over36}\right)\ep^4\right)\nonumber\\
&&+k^2\left(1\mp\ep^2+\ep^4\right)
\eea
where we again use $\gam$ from eq.~(\ref{gamdef}) with $\ex=2$; so $\gam=3$ for $\delta\phi_\parallel$, and $\gam=1$ for $\delta\phi_\perp$. 

We then substitute into the harmonic expansion (\ref{harmonics}). Previously we studied the first instability band by tracking the leading odd harmonics $\omega=-m,+m$. To study the second instability band we need to look at the leading even harmonics $\omega=-2m,0,+2m$. 
The $\omega=0$ mode is easily solved for, leaving just two unknown coefficients. 
We note that these couple to $\omega=-4m,+4m$, giving finite corrections. However, a reasonable approximation for the central results, arises from just focussing on the $\omega=-2m,+2m$ harmonics. We find that the $2\times2$ matrix problem for the second instability band takes the form
\bea
&&{d\over d\tau}\left(\!\begin{array}{c}
\delta\phi_{+2m}\\
\delta\phi_{-2m}
\end{array}\!\right)\nonumber\\
&&={i\over4m}\!\left(\!\begin{array}{cc}
\MA-4m^2-{\Mq^2\over \MA} & \Ml-{\Mq^2\over\MA}\\
{\Mq^2\over\MA}-\Ml & 4m^2+{\Mq^2\over \MA}-\MA
\end{array}\!\right)\!
\left(\!\begin{array}{c}
\delta\phi_{+2m}\\
\delta\phi_{-2m}
\end{array}\!\right)\,\,\,\,\,\,\,\,\,\,\,\,\,
\eea
The eigenvalues of this matrix are the Floquet exponents of the second band
\beq
\mu_k = {1\over4m}\sqrt{\left(\Ml - {\Mq^2\over\MA}\right)^2-\left(\MA-4m^2-{\Mq^2\over\MA}\right)^2}
\eeq
The full expression for $\mu_k$ after substituting in the above values for $\MA,\,\Mq,\,\Ml$ is somewhat complicated, but it suffices to discuss its features. 

Firstly, lets discuss at what $k$ value the band starts at in the small amplitude limit. If we take $\phi_a\to0$, then $\Mq\to0$, $\Ml\to0$, and $\MA\to k^2+m^2$, and $\mu_k$ becomes
\beq
\mu_k\to \pm{i\over 4m}(k^2-3m^2)
\eeq
Hence the value of $k_*$ that sets the imaginary part of $\mu_k$ to zero, and hence corresponds to the start of the instability band, is 
\beq
k_*=\sqrt{3}\,m
\eeq
In Section \ref{Quantum} we will explain this wavenumber as arising from $4\phi\to2\phi$ particle annihilations.

Next, we discuss the shape and width of the instability band. If we work only to $\mathcal{O}(\ep^2)$ we find that both the left hand and right hand edge of the instability band coincide. As a function of amplitude this merely provides the overall {\em bending} of the band. If we express this in terms of the amplitude $\phi_a$, we find that to $\mathcal{O}(\ep^2)$ we have
\beq
k_{edge} = \sqrt{3}\,m+{(6-\gam)\over4\sqrt{3}}{\lambda\,\phi_a^2\over m}
\label{bend}\eeq
On the other hand, we find a splitting between the left and right hand edges at $\mathcal{O}(\ep^4)$. This splitting, which gives the width of the band, is found to be
\beq
\Delta k = k_{r,edge}- k_{l,edge} = {\gam|\gam-1|\over 64\sqrt{3}}{\lambda^2\phi_a^4\over m^2}
\label{bandwidth}\eeq
So for the isocurvature modes ($\gam=1$) the width is zero. This means there is no instability band. This is in accord with the bottom panel of Fig.~\ref{4PanelView}. On the other hand, for the adiabatic modes ($\gam=3$) there does exist a finite width and hence a narrow instability band. From eq.~(\ref{bend}) we see that this band bends to the right (higher $k$) for $\lambda>0$, or bends to the left (lower $k$) for $\lambda<0$, as we increase the amplitude. This explains the features seen in the narrow instability band in the upper panel of Fig.~\ref{4PanelView}. 
Since this band is due to $4\phi\to2\phi$ annihilations, as we explain in Section \ref{Quantum}, it requires energy density perturbations, and so it makes sense that it does not exist for the isocurvature modes.

\subsection{Negative Mass Squared} \label{NegMass}

We now consider the case of negative mass squared ($m^2<0$ and $\lambda>0$) at small amplitudes. So we now study a Higgs-type of potential and expand around the true vacuum $|\vec\phi|=\phi_{vev}=|m|/\sqrt{\lambda}$. We put the VEV in the $\N^{th}$ direction in field space and write the field as
\beq
\vec\phi = \{\phi_{1},\ldots,\phi_{\N-1},{|m|\over\sqrt{\lambda}}+\sig\}
\eeq
The dimension 4 potential becomes
\bea
V  &=& {1\over2}m_\sig^2\sig^2+{\lambda_3\over3}\sig^3+{\lambda\over4}\sig^4\nonumber\\
&&+{\lambda_3\over3}\sig\sum_i\phi_i^2+{\lambda\over2}\sig^2\sum_i\phi_i^2+{\lambda\over4}\Big{(}\sum_i\phi_i^2\Big{)}^2\,\,\,\,\,\,\,\,\,\,\,
\eea
where the sum over $i$ is from 1 to $\N-1$ of the Goldstones $\phi_i$. The mass of the $\sig$ field and the cubic coupling are
\bea
m_\sig &=& \sqrt{2}\,|m|\label{msig}\\
\lambda_3 &=& 3\sqrt{\lambda}\,|m|\label{lam3}
\eea

Let us discuss radial motion of the background described by the field $\sig_0$. This satisfies the equation of motion
\beq
\ddot\sig_0+m_\sig^2\,\sig_0+\lambda_3\,\sig_0^2+\lambda\,\sig_0^3=0
\eeq
We again use a small $\ep$ expansion to solve this to leading order. Due to the cubic interaction, there will be both odd and even harmonics in the expansion. We will not go through the full details, but we find that to leading order, the solution is
\beq
\sig_0(\tau) = \ep{2\,m_\sig\over\sqrt{3|\lambda-10\,\lambda_3^2/(9\,m_\sig^2)|}}\cos(m_\sig\,\tau)
\eeq 
This result is general for any choice of $\lambda$ and $\lambda_3$. Of course by using the relationships in eqs.~(\ref{msig},\,\ref{lam3}) the denominator can be simplified. Compared to the analysis of Section \ref{First}, where there was no cubic term, we see that in some sense the ``effective" $\lambda$ has been shifted to
\beq
\lambda\to\lambda-{10\,\lambda_3^2\over9\,m_\sig^2} = -4\,\lambda
\label{lambdashift}\eeq
With this understanding we can immediately use eq.~(\ref{muSA}) (with $\ex=2$) to write down the answer for the Floquet exponent in the first instability band for the parallel, or adiabatic, perturbations. We find
\beq
\mu_k = {k\over2\,m_\sig}\sqrt{6\,\lambda\,\sig_a^2-k^2}
\eeq
We now see opposite behavior of the adiabatic mode for $m^2<0$ compared to $m^2>0$. Now we see that for $\lambda>0$ there is an instability band for small $k$. On the other hand, when $m^2>0$ we only saw such a band for $\lambda<0$. This makes sense from the point of view of the pressure analysis of Part 1 \cite{Part1}. Indeed one can check that the cubic term induces a {\em negative} pressure for small amplitudes, even though the quartic term is positive. For sufficiently large amplitudes, the pressure returns to being positive, as the positive quartic term dominates, and the band is shut off for small $k$. This explains the change between Figs.~\ref{4PanelView} and \ref{NegativeMass}. In the $\lambda>0$, $m^2>0$ plot for the parallel perturbations we saw that a thin band started at $k=\sqrt{3}\,m$. In the $m^2<0$ plot for the parallel perturbations we saw this band thicken and extend all the way down to $k=0$, in agreement with our new analysis.

For the orthogonal, or isocurvature, perturbations the equation of motion for small amplitudes is
\beq
\ddot{\!\delta\phi}_{\perp i}+(k^2+\sqrt{2\lambda}\,m_\sig\,\sig_a\cos(m_\sig t))\delta\phi_{\perp i} = 0
\eeq
In fact we will not require the distinction between $\tau$ and $t$ for this leading order analysis, so we have written the argument of the cosine as $m_\sig t$.
Since this is of the form of the Mathieu equation, we can follow through the steps of Section \ref{First} to readily obtain the Floquet exponent. We find
\beq
\mu_k = {1\over2}\sqrt{8\,\lambda\,\sig_a^2-\left({4k^2\over m_\sig}-m_\sig\right)^2}
\eeq
We now see that this band no longer begins at $k=0$, instead it begins at
\beq
k_* = {m_\sig\over2} = {|m|\over\sqrt{2}}
\eeq
This is precisely what is observed in the lower panel of Fig.~\ref{NegativeMass}. The reason the band does not exist at small $k$ is because of the complementary behavior to the adiabatic mode. Since the ``ordinary" pressure is negative due to the cubic term rendering the adiabatic mode unstable for small $k$, the ``auxiliary" pressure is positive rendering the isocurvature modes stable for small $k$. Nevertheless there is a thick instability band beginning at $k=m_\sig/2$; we shall explain its origin as the decay of the ``Higgs" field into a pair of Goldstones $\phi_i$ in the next Section.

\section{Many Particle Quantum Mechanics} \label{Quantum}

In the previous sections we have seen several interesting results, including physical explanations in terms of pressure, small amplitude analysis, etc.
Some of the salient results are (sharply true at small amplitudes): 
\begin{enumerate}[(i)]
\item At small $k$ and $m^2>0$, the adiabatic mode is unstable for $\lambda<0$, and the isocurvature modes are unstable for $\lambda>0$.
\item At small $k$ and $m^2<0$, the adiabatic mode is unstable (we require $\lambda>0$ here). 
\item At $k\sim\sqrt{3}\,m$ and $m^2>0$, the adiabatic mode is unstable (for either sign of $\lambda$). 
\item At $k\sim|m|/\sqrt{2}$ and $m^2<0$, the isocurvature modes are unstable.
\end{enumerate}
In this Section we will give the underlying quantum mechanical explanation for each one of these facts. We will particularly emphasize point (i) which most clearly highlights the complementary behavior between adiabatic and isocurvature modes; their relative stability is determined by the sign of $\lambda$.

The reason there should be a quantum mechanical explanation is that underlying the field theory should be a more fundamental description in terms of many quantum particles. Indeed the above classical scalar field theory analysis is only a good approximation if it approximates the behavior of some kind of condensate of scalar bosons.

\subsection{Nonrelativistic Theory}

\begin{figure}[t]
\includegraphics[width=0.75\cw]{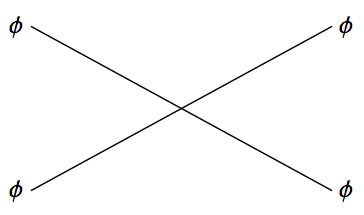}\\\bigskip
\includegraphics[width=0.75\cw]{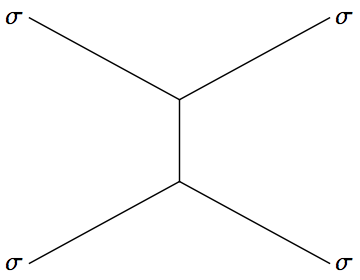}
\caption{Representative Feynman diagrams of two important nonrelativistic processes. Upper panel: $2\phi\to2\phi$ scattering from 4-point vertex. Lower panel: $2\sig\to2\sig$ scattering from 3-point vertex (relevant for Higgs potential).}
\label{4Vertex}\end{figure}

Let us begin by focussing on dimension 4 potentials with $m^2>0$ and $\lambda\neq0$. For $k\ll m$ we should be able to use the nonrelativistic treatment of a collection of massive scalars. In the case of a complex field, there are 2 kinds of identical species: particles $\phi$ and antiparticles $\bar{\phi}$. In the nonrelativistic regime these particles only interact with one another via $2\phi\to2\phi$ scattering from a 4-point vertex; see upper panel of Fig.~\ref{4Vertex}. 
The associated matrix element is a constant, namely $\mathcal{M}=-3\,i\,\lambda$. By Fourier transforming, we obtain the following 2-body potential
\beq
V({\bf x}_1 - {\bf x}_2) = {3\,\lambda\over 4\,m^2}\,\delta({\bf x}_1 - {\bf x}_2)
\eeq
where ${\bf x}_1$ and ${\bf x}_2$ are the positions of a pair of particles/antiparticles. By considering $N_\phi$ particles and $N_{\bar{\phi}}$ antiparticles and summing, we obtain the the following quantum Hamiltonian
\bea
\hat{H} &=& \sum_a^{\Num_\phi}{\hat{p}_a^2\over 2m}+\sum_{\bar{a}}^{\Num_{\bar{\phi}}}{\hat{p}_{\bar{a}}^2\over 2m}+\sum_{a<b}^{\Num_\phi,\Num_\phi}V(\hat{\bf x}_a-\hat{\bf x}_b)\nonumber\\
&&+\sum_{a<\bar{b}}^{\Num_\phi,\Num_{\bar{\phi}}}V(\hat{\bf x}_a-\hat{\bf x}_{\bar{b}})+\sum_{\bar{a}<\bar{b}}^{\Num_{\bar{\phi}},\Num_{\bar{\phi}}}V(\hat{\bf x}_{\bar{a}}-\hat{\bf x}_{\bar{b}})
\eea
where we have indicated particles by index $a$ or $b$, and antiparticles by index $\bar{a}$ or $\bar{b}$.
So a positive $\lambda$ implies a {\em repulsive} force between the particles (and antiparticles), while a negative $\lambda$ implies an {\em attractive} force between the particles (and antiparticles). In fact this is also true if we went beyond the quartic interaction $\sim \lambda\,\phi^4$ to a general potential 
$\sim\lambda\,\phi^{2\ex}$ with $\ex\ge2$ an integer. Say for $\ex=3$, we have a 3-body contact interaction, whose attraction/repulsion is determined by the sign of $\lambda$.

\begin{figure}[t]
\includegraphics[width=0.9\cw]{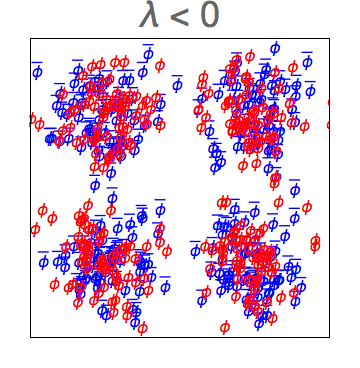}\\
\includegraphics[width=0.9\cw]{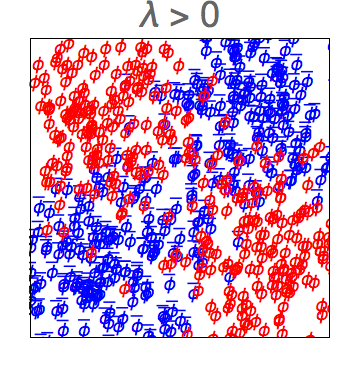}
\caption{A cartoon of the behavior of (nonrelativistic) particles $\phi$ in red and antiparticles $\bar\phi$ in blue. 
In the upper panel, $\lambda<0$, we depict both particles and antiparticles attracting with statistically uniform number density ($\nd=\nd_{\phi}-\nd_{\bar{\phi}}$). In the lower panel, $\lambda>0$, we depict particles clumping with particles and antiparticles clumping with antiparticles with statistically uniform energy density ($\ed=m(\nd_\phi+\nd_{\bar{\phi}})$ in nonrelativistic limit).}
\label{Cartoon}\end{figure}

Lets imagine an initial homogeneous configuration of equal numbers of particles and antiparticles. Indeed for a classical background that evolves {\em radially} in field space, the background number density of particles minus antiparticles is zero. 
If $\lambda>0$, the particles will want to remain evenly distributed due to their mutual repulsion. On the other hand, if $\lambda<0$, the particles will want to clump together under their mutual attraction. A cartoon of this behavior is depicted in upper panel of Fig.~\ref{Cartoon}. 
(In the figure we drew particles and antiparticles from Gaussians centered around each of the 4 quadrants to illustrate this.)
This is the physical explanation as to why the adiabatic mode is stable with $\lambda>0$ and unstable with $\lambda<0$ for small $k$. At the nonlinear level, this can produce localized structures known as oscillons \cite{Amin:2010dc,Gleiser:2011xj,Amin:2011hj,Lozanov:2014zfa}, which eventually annihilate away \cite{Hertzberg:2010yz}. When there is an overabundance of particles to anti-particles (see Section \ref{Baryogenesis}) this can produce stable objects known as Q-balls \cite{Coleman:1985ki}.

But what about the isocurvature mode? If $\lambda>0$ the particles will try to repel each other and remain with a homogeneous energy density, with $\ed$ $=m(\nd_\phi+\nd_{\bar{\phi}})$ in nonrelativistic limit. However there are different types of homogeneous configurations, since we have 2 species available. For instance, the particles and the antiparticles can remain homogeneous. Or the particles can move to certain regions in space, while the antiparticles move to other regions of space, in such a way that the total energy density remains constant. This latter arrangement involves a change to $\nd=\nd_\phi-\nd_{\bar{\phi}}$, the local number density of particles minus antiparticles; an isocurvature mode. Either arrangement seems to minimize the energy. However, it is the latter arrangement that is favored due to Bose-Einstein statistics, which favors particles clumping with particles, and antiparticles clumping with antiparticles. A cartoon of this behavior is depicted in lower panel of Fig.~\ref{Cartoon}. (In the figure we drew particles and antiparticles from Gaussians centered at opposite quadrants to illustrate this.) This explains why the isocurvature mode is unstable when $\lambda>0$ for small $k$. These last pair of arguments explain result (i) and the low $k$ region of Fig.~\ref{4PanelView}.

For $m^2<0$ and $\lambda>0$ we can expand around the vacuum expectation value for the field. This induces a cubic interaction for the Higgs field $\sig$, as we explored in Section \ref{NegMass}. This 3-point interaction alters the $2\to2$ scattering, as given by the Feynman diagram in the lower panel of Fig.~\ref{4Vertex}. The adiabatic modes carry a mass (the ``Higgs" particles) and so we can take the nonrelativistic limit. In the nonrelativistic limit these matrix elements can be computed and Fourier transformed. The result is again a delta-function potential with the coupling altered as $\lambda\to-4\,\lambda$, precisely as we saw earlier in eq.~(\ref{lambdashift}). This switches the sign of the potential. For $\lambda>0$, the 2-body potential is now attractive, due to the 3-body $\sig$ exchanges. Hence it makes good sense that the adiabatic mode is now unstable. This argument explains result (ii) and the low $k$ region of Fig.~\ref{NegativeMass}.

\subsection{Relativistic Theory}

\begin{figure}[t]
  \includegraphics[width=0.75\cw]{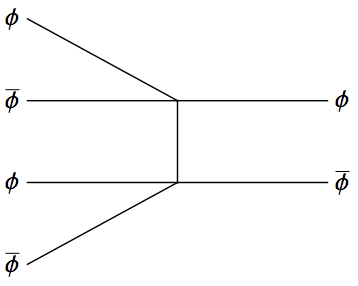}\\\bigskip
   \includegraphics[width=0.75\cw]{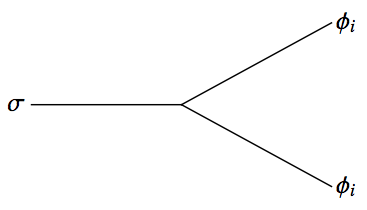}
  \caption{Representative Feynman diagrams of two important relativistic processes. Upper panel: $2\phi+2\bar\phi\to\phi+\bar{\phi}$ annihilation. Lower panel: $\sig\to\phi_i+\phi_i$ decay (relevant for Higgs potential).}
\label{Annihilations}\end{figure}

Returning again to a regular mass term with a quartic interaction, we now allow for relativistic processes to occur. 
Since the homogeneous background is a dense condensate of bosons, a quartic interaction can lead to annihilations. Due to the conserved particle number, we are allowed to have processes, such as $2\phi+2\bar\phi\to\phi+\bar{\phi}$. This is the leading annihilation process, whose corresponding Feynman diagram is given in the upper panel of Fig.~\ref{Annihilations}. 
Since the annihilating particles are homogeneously distributed, they can be taken to be nonrelativistic. The effective mass per particle is $m$ in the small amplitude limit. So the kinematics of this process is
\beq
4\,m=2\sqrt{m^2+k_*^2} \implies k_*=\sqrt{3}\,m
\eeq
Notice that since this process occurs simply due to kinematics, it will occur regardless of the sign of $\lambda$. The only significance of $\lambda$ is that it alters the effective mass of the homogeneous clump at finite amplitudes. So for $\lambda>0$ the effective mass of the annihilating particles is raised, causing the corresponding outgoing $k$ to be raised. The opposite is true for $\lambda<0$. We also note that a process such as this involves a redistribution of the local energy density, and hence it is an adiabatic mode. These arguments explain result (iii) and the thin bands of Fig.~\ref{4PanelView}.

Finally we consider the Higgs type of potential, and consider relativistic processes. Since the $\phi_i$ ($i=1,\ldots,\N-1$) are massless, a basic decay process can take place between the background Higgs field and the daughter Goldstones $\sig\to\phi_i+\phi_i$. This process is given in the lower panel of Fig.~\ref{Annihilations}. The kinematics of this process is
\beq
m_\sig = 2\,k_* \implies k_*={|m|\over\sqrt{2}}
\eeq
Since the phase space of one particle decay is much greater than the phase space of 4 particle annihilations, this band should be much thicker than the band described above in the unbroken theory. We also mention that other annihilations are presumably allowed in the broken theory, such as $\sig+\sig\to\phi_i+\phi_i$. However this will still result in a rather thin band, beginning at $k_*=\sqrt{2}\,|m|$, and beyond the regime plotted in Fig.~\ref{NegativeMass}. Altogether these arguments explain result (iv) and the thick band in the lower panel of Fig.~\ref{NegativeMass}.

\section{Perturbation Growth with Hubble Expansion} \label{Hubble}

Earlier when we computed the growth of fluctuations, we ignored the effects of Hubble expansion. In this Section we would like to reinstate the effects of expansion. Accordingly, we return to the important case in which the background field evolves radially in field space, as initially established by inflation.

The expansion causes a breaking of the periodicity of the pump $h(t)$ as the background amplitude and perturbation wavenumber redshift.
So, strictly speaking, this means that we can no longer use Floquet theory. Instead we can numerically solve the linearized equations to capture the evolution; which we will do shortly in Section \ref{HubbleNumerical}. On the other hand, the Floquet theory is still very useful to provide qualitative and semi-quantitative results, as we now explain.

\begin{figure*}[t]
  \includegraphics[width=\cw]{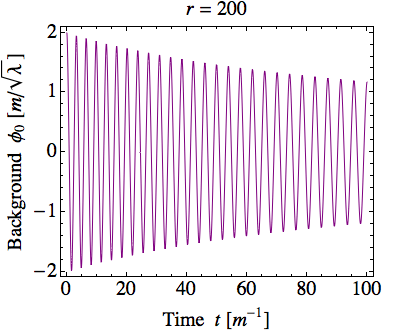}
  \includegraphics[width=\cw]{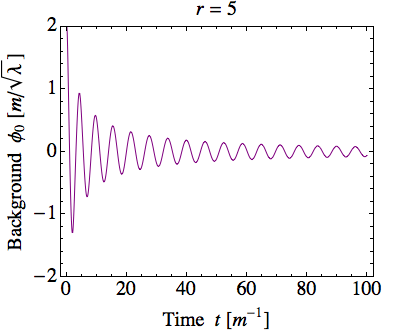}
\caption{Evolution of background $\phi_0$ with Hubble expansion after inflation for dimension 4 potentials, with $m^2>0$ and $\lambda>0$. 
In the left panel $\Pv\equiv\sqrt{\lambda}\,\mpl/m=200$ (slower redshifting). 
In the right panel $\Pv\equiv\sqrt{\lambda}\,\mpl/m=5$ (faster redshifting). 
We have plotted $\phi_0$ in units of $m/\sqrt{\lambda}$ and $t$ in units of $1/m$. At somewhat earlier times, the field slow-rolls during inflation.}
\label{HubbleBackground}\end{figure*}

\subsection{Slow Redshift Approximation}

If we continue to study modes that are sub-horizon, then we can introduce an approximate treatment of the expansion.
The idea is that on the scales of interest, the background changes only slightly over a periodic of oscillation. So this presents a type of slow, or ``adiabatic", approximation wherein we can utilize Floquet theory on short time scales, as well as accounting for the slow redshifting in an approximate fashion (note that the use of the word ``adiabatic" here does {\em not} refer to the adiabatic modes discussed elsewhere in this paper).

The first alteration is to take the Floquet exponent from the previous section $\mu_k$ and replace its argument by the physical wavenumber $k_p$
\beq
k\to \kp(t) = {k\over a(t)},\,\,\,\,\,\mu_k\to\mu_{\kp}
\label{Physk}\eeq
as it is this quantity that appears in the equation of motion. Secondly, we note that Hubble introduces a redshifting in the amplitude of the pump. The details depend on the choice of the potential $V$. We can summarize this as an effective evolution in the amplitude as
\beq
\phi_a\to\phi_a(t)
\eeq
which we take to be slowly varying. There is also an overall rapid oscillation in the perturbations, captured by some quasi-periodic function $f(t)$. 
At late times this is roughly $f(t)\approx\cos(m\,t+\theta)$, where $\theta$ is some phase (see ahead to Fig.~\ref{HubbleBackground}).

Thirdly, since the Floquet exponent is now time dependent due to redshifting, the growth is altered. The adiabatic approximation is to replace the exponential growth by an integral as follows (we denote the real part of the Floquet exponent by $\mu_{k_p}$ here)
\beq
\exp[\mu_k\,(t-t_i)]\to \exp\left[\int^t_{t_i} dt'\,\mu_{\kp}(t')\right]
\eeq
where $t_i$ is some initial time (say, the end of inflation).
Now it is convenient to use the chain rule to recast the integral over time as an integral over physical wavenumber, as this is what the Floquet exponent explicitly depends on. Using (\ref{Physk}) we can express this as
\beq
\int^t_{t_i} dt'\,\mu_{\kp}(t') = \int_{k_{p}(t)}^{k_{if}} \!d\ln k_p\left(\mu_{k_p}\over H\right)
\eeq
where the integrand is evaluated along the appropriate curve in the Floquet chart (see ahead to the dashed green curve in Fig.~\ref{Zoom}).
Note the limits of integration: We have placed the late time wavenumber $k_p(t)$ at the lower endpoint, and the initial wavenumber $k_{pi}$ at the upper endpoint, so the integral is ordered in a standard way.

So altogether, a rough inclusion of Hubble expansion is to write the field fluctuations as
\beq
{\delta\phi(k,t)\over\delta\phi(k,t_i)}\sim\left(\phi_a(t)f(t)\over\phi_a(t_i)\right) \exp\left[\int_{k_{p}(t)}^{k_{if}} \!d\ln k_p\left(\mu_{k_p}\over H\right)\right]
\eeq
where $\mu_{\kp}$ is the (real part of) Floquet exponent for either the adiabatic or isocurvature modes, as appropriate. 

For dimension 4 potentials, it is convenient to introduce the parameter $\Pv\equiv\sqrt{|\lambda|}\,\mpl/|m|$, as mentioned earlier. 
The reason this parameter is useful is that the combination $\mu_{k_p}/(\Pv H)$ is independent of parameters at a fixed amplitude and wavenumber, as reported on in our earlier Floquet charts. The exponent can be written as
\beq
\exp\left[\Pv\!\int_{k_{p}}^{k_{if}} \!d\ln k_p\left(\mu_{k_p}\over H\Pv\right)\right]
\eeq
which shows that the growth is approximately exponential in $\Pv$. We say ``approximately" because the details of the motion through the band has some parameter dependence, though it is relatively small.

\subsection{Numerical Results for Growth}\label{HubbleNumerical}

We have numerically solved for the background field $\phi_0(t)$, allowing for Hubble expansion, for different values of $\Pv$. For $m^2>0$ and $\lambda>0$ the result is given in Fig.~\ref{HubbleBackground}. In the left panel we have taken $\Pv=200$ and in the right panel we have taken $\Pv=5$. As the plots show, for higher values of $\Pv$ the redshifting is slow and for lower values of $\Pv$ the redshifting is fast. So in the former case, the background $\phi_0(t)$ is almost periodic on the time scale of a small number of oscillations. This means that Floquet theory provides a good approximation to the behavior, as described in the previous subsection. In the latter case, the background $\phi_0(t)$ changes rather significantly from one cycle to the next, so the Floquet theory becomes less accurate. 

\begin{figure*}[t]
  \includegraphics[width=1.04\cw]{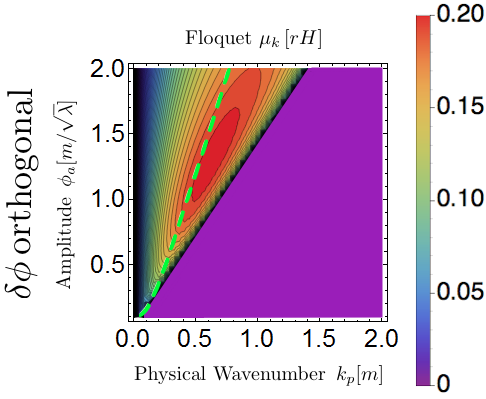}\,\,\,
  \includegraphics[width=\cw,height=\htn]{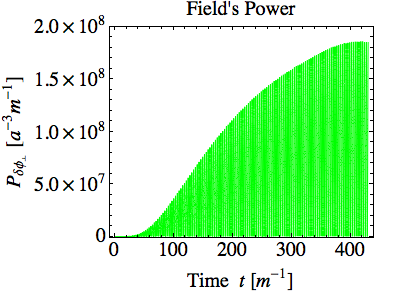}
  \caption{Left panel: Contour plot of the real part of Floquet exponent $\mu_k$ for dimension 4 potentials as a function of physical wavenumber $k_p$ and background amplitude $\phi_a$ for $\delta\phi_{\perp}$, with $m^2>0$ and $\lambda>0$ (zoomed in region of Fig.~\ref{4PanelView}'s lower left panel). The dashed green line indicates the redshifting physical wavenumber and amplitude of a fixed comoving wavenumber $k=0.4\,a_0\,m$, where $a_0$ is the scale factor at $\phi_a=m/\sqrt{\lambda}$. We have plotted $\mu_k$ in units of $\Pv H$ where $\Pv\equiv \sqrt{\lambda}\,\mpl/m$, $k_p$ in units of $m$, and $\phi_a$ in units of  $m/\sqrt{\lambda}$. Right panel: Growth in power $P_{\delta\phi_\perp}$ as a function of time for the same comoving wavenumber $k=0.4\,a_0\,m$ with $\Pv=20$. We have plotted $P_{\delta\phi_\perp}$ in units of $1/(a^3\,m)$ and $t$ in units of $1/m$. For $\lambda<0$ (not shown here) qualitatively similar behavior occurs for $\delta\phi_\parallel$.}
\label{Zoom}\end{figure*}

It is, however, the case of large $\Pv$ that is of most interest from the point of view of self-resonance. In Fig.~\ref{Zoom} left panel we have plotted the behavior of a redshifting physical wavenumber and amplitude for a fixed comoving wavenumber; $k=0.4\,a_0\,m$, where $a_0$ is the scale factor at $\phi_a=m/\sqrt{\lambda}$. We have chosen this as a representative wavenumber that passes through the central instability band. We have chosen $m^2>0$ and $\lambda>0$ here and shown the instability associated with the isocurvature modes. For $\lambda<0$ (not shown here) we find the behavior to be qualitatively similar for the adiabatic mode.

In the next Section we will discuss the power spectra of field fluctuations $P_{\delta\phi}(k,t)$, which is related to the square of the fluctuations $\delta\phi(t)$. In Fig.~\ref{Zoom} right panel we have plotted the evolution of the power spectra, accounting for Hubble expansion, for the same comoving wavenumber $k=0.4\,a_0\,m$. We see that the growth appears exponential, but is reduced at late times as the redshift decreases the value of $\mu_{k_p}/(\Pv H)$ towards zero, and so its amplitude asymptotes to a constant. There are of course rapid oscillations on top of this.

\section{Distribution of Densities $\delta\ed,\,\delta\nd$} \label{PositionSpace}

In this section we discuss the quantization of the perturbations $\delta\phi$ and their initial conditions, and present results for the distribution of fluctuations in $k$ space and position space. We continue to account for Hubble expansion where necessary.

\subsection{Quantization of Perturbations} \label{QuantizePert}

When the background $\vec\phi_0$ evolves radially in field space, as preferred by inflation, the perturbations $\delta\phi_\parallel$ and $\delta\phi_\perp$ are decoupled at linear order, as discussed in Section \ref{Linearized}. 

Since these modes are decoupled, they can be readily quantized in the Heisenberg picture. We write the operator for parallel fluctuations as
\beq
\hat{\delta\phi}_\parallel({\bf x},t) = \!\int\!{d^3k\over(2\pi)^3}\!\left[v_{\parallel,k}(t)\,\hat{a}_k\,e^{i{\bf k}\cdot{\bf x}}+v_{\parallel,k}^*(t)\,\hat{a}_k^\dagger\,e^{-i{\bf k}\cdot{\bf x}}\right]\,\,\,
\eeq
and the operator for orthogonal fluctuations as
\beq
\hat{\delta\phi}_\perp({\bf x},t) = \!\int\!{d^3k\over(2\pi)^3}\!\left[v_{\perp,k}(t)\,\hat{b}_k\,e^{i{\bf k}\cdot{\bf x}}+v_{\perp,k}^*(t)\,\hat{b}_k^\dagger\,e^{-i{\bf k}\cdot{\bf x}}\right]\,\,
\eeq
where $v_\parallel$ and $v_\perp$ are the respective mode functions. The creation and annihilation operators satisfy the standard quantization condition
\bea
\left[\hat{a}_k,\hat{a}_{k'}^\dagger\right]&=&(2\pi)^3\delta^3({\bf k}-{\bf k}') \\
\left[\hat{b}_k,\hat{b}_{k'}^\dagger\right]&=&(2\pi)^3\delta^3({\bf k}-{\bf k}') 
\eea
As this is a free theory, the mode functions satisfy the classical equations of motion that we previously discussed for $\delta\phi$ in eqs.~(\ref{PertEqn1},\,\ref{PertEqn2}). So, allowing for Hubble expansion, we have
\bea
&&\ddot v_\parallel+3H\dot v_\parallel+\left({k^2\over a^2}+V''(\phi_0)\right)v_\parallel=0\label{vEqn1}\\
&&\ddot v_{\perp}+3H\dot v_{\perp}+\left({k^2\over a^2}+{V'(\phi_0)\over\phi_0}\right)v_{\perp}=0\label{vEqn2}
\eea
We assume that at early times, the mode functions are in their Minkowski vacua, and then evolve; this is the Bunch-Davies vacuum. So at early times we require the initial condition
\beq
v_{\parallel,k},\,v_{\perp,k} \to {e^{-i\omega_k t}\over\sqrt{2\,\omega_k\,a^3}}
\eeq
where the frequency is $\omega_k=\sqrt{m^2+k^2/a^2}$. In fact at early times it is sufficient to use $\omega_k\to k/a$.

We can go further and quantize the perturbations in energy density $\delta\ed$ and number density $\delta\nd$. We write the operator for energy density  (adiabatic) fluctuations as
\beq
\hat{\delta\ed}({\bf x},t) = \!\int\!{d^3k\over(2\pi)^3}\!\left[\emo_{k}(t)\,\hat{a}_k\,e^{i{\bf k}\cdot{\bf x}}+\emo_{k}^*(t)\,\hat{a}_k^\dagger\,e^{-i{\bf k}\cdot{\bf x}}\right]\,\,\,
\eeq
and the operator for number density (isocurvature) fluctuations as
\beq
\hat{\delta\nd}({\bf x},t) = \!\int\!{d^3k\over(2\pi)^3}\!\left[\nmo_{k}(t)\,\hat{b}_k\,e^{i{\bf k}\cdot{\bf x}}+\nmo_{k}^*(t)\,\hat{b}_k^\dagger\,e^{-i{\bf k}\cdot{\bf x}}\right]\,\,
\eeq
We can relate these density fluctuations to the field fluctuations by using the quantized versions of the linearized energy density and number densities, as we defined in Part 1 \cite{Part1}. 
This leads to the following relationship between the density mode functions $\emo$ and $\nmo$ and the field mode functions $v_\parallel$ and $v_\perp$
\bea
\emo_k(t)&=&\left(\dot\phi_0{\partial\over\partial t}+V'(\phi_0)\right)v_{\parallel,k}(t)\\
\nmo_k(t)&=&-\left(\phi_0{\partial\over\partial t}-\dot\phi_0\right)v_{\perp,k}(t)
\eea
In the sub-horizon limit, the Floquet theory, pressure analysis, and so on, of the earlier sections, determines these mode functions.

\subsection{Probability Distributions}

So we have the following physical variables of interest
\beq
\Var = \{\delta\phi_\parallel,\,\delta\ed,\,\delta\phi_\perp,\,\delta\nd\}
\eeq
where the first pair are independent of the last pair.
By placing the field fluctuations in their ground state, the wave-functional $\Psi$ for each variable is a Gaussian. In the Schr\"odinger picture, the corresponding probability distribution for any of these variables is
\bea
\mathcal{P}[\Var,t]  
\propto \exp\!\left[-{1\over2}\int\!{d^3k\over(2\pi)^3}{|\Var_k|^2\over P_\Var(k,t)}\right]\,\,\,
\eea
where $P_\Var$ is the power spectrum for each variable, defined through the equal time 2-point correlation function as
\beq
\langle \hat\Var_k(t)\,\hat\Var_{k'}(t) \rangle = (2\pi)^3 \delta^3({\bf k}+{\bf k}')\,P_\Var(k,t)
\eeq
Since the background breaks the time translation symmetry, the power spectra depend on time, as we have indicated.
It is straightforward to show that they are given by the square of their corresponding mode functions
\bea
P_{\delta\phi_\parallel}(k,t) &=& |v_{\parallel,k}(t)|^2,\,\,\,\,P_{\delta\ed}(k,t) = |\emo_{k}(t)|^2\,\,\,\,\,\,\,\\
P_{\delta\phi_\perp}\!(k,t) &=&  |v_{\perp,k}(t)|^2,\,\,\,P_{\delta\nd}\!(k,t) =  |\nmo_{k}(t)|^2
\eea
This furnishes the probability distribution for the fields. 

We would also like to have the probability distributions for their time derivatives. So lets define the variables
\beq
\Pi = \{\dot{\delta\phi}_\parallel,\,\dot{\delta\ed},\,\dot{\delta\phi}_\perp,\,\dot{\delta\nd}\}
\eeq
We then also have a Gaussian distribution for these ``momenta" (these are not canonically normalized momenta, as this would require the inclusion of additional powers of the scale factor, etc)
\bea
\mathcal{P}[\Pi,t]  
\propto \exp\!\left[-{1\over2}\int\!{d^3k\over(2\pi)^3}{|\Pi_k|^2\over P_\Pi(k,t)}\right]\,\,\,
\eea
where the power spectra for the momenta are given as the square of the time derivatives of the corresponding mode functions.

In order to simulate the fields, we can use these probability distributions to draw the fields at a given moment in time. Of course quantum mechanically, we cannot specify both $\Var$ and $\Pi$ simultaneously as they do not commute with one another (uncertainty principle). On the other hand, in order to provide the initial conditions of a classical simulation, one can draw from both and then evolve under the classical equations of motion. This naturally loses the non-commutativity of the field and its momentum conjugate, as is required for a classical simulation. Interestingly, by drawing from both at an initial time, then evolving under the classical equations of motion, then repeating and ensemble averaging, one actually reproduces the correct quantum expectation values in the linear approximation.

\begin{figure}[t]
\hspace{0.6cm}  \includegraphics[width=0.9\cw]{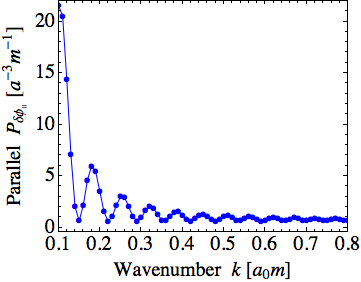}\vspace{0.5cm}
  \includegraphics[width=\cw]{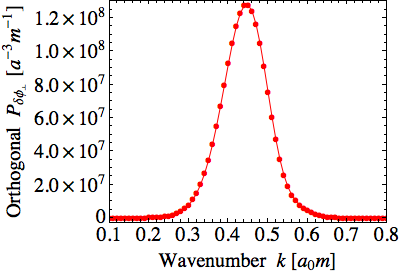}
  \caption{Late time power spectra $P_{\delta\phi}$ of field fluctuations as a function of comoving wavenumber $k$ for dimension 4 potentials, with $m^2>0$, $\lambda>0$, and $\Pv=20$.
In the upper panel are (non-resonant) parallel fluctuations $P_{\delta\phi_\parallel}$.
In the lower panel are (resonant) orthogonal fluctuations $P_{\delta\phi_\perp}$.
We have plotted $P_{\delta\phi}$ in units of $1/(a^3\,m)$ and comoving $k$ in units of $a_0\,m$, where $a_0$ is the scale factor at $\phi_a=m/\sqrt{\lambda}$. For $\lambda<0$ (not shown here) the resonances are interchanged.}
\label{Power}\end{figure}

\begin{figure*}[t]
  \includegraphics[width=0.98\cw]{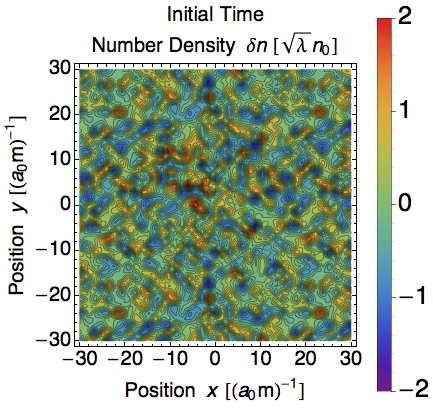}\,
  \includegraphics[width=1.07\cw]{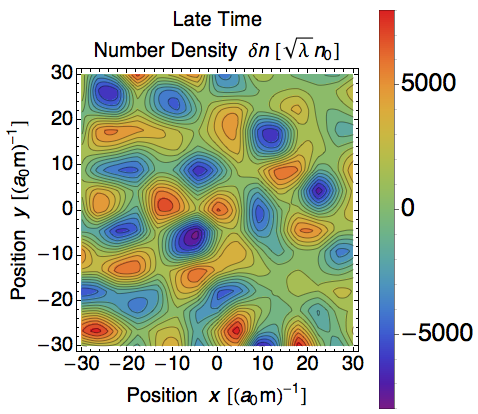}
\caption{Two-dimensional slice of number density fluctuations $\delta\nd$ in position space for dimension 4 potentials, with $m^2>0$, $\lambda>0$, and $\Pv=20$. The left panel is at an initial time, which is dominated by UV modes (cutoff at $k_{UV}=3\,a_0\,m$), associated with {\em virtual} particles-antiparticles. The right panel is at a late time, which is dominated by IR modes that have exponentially grown, associated with separate regions of {\em real} particles and antiparticles. We have plotted $\delta\nd$ in units of $\sqrt{\lambda}\,\nd_0$ and comoving position in units of $1/(a_0\,m)$. For $\lambda<0$ (not shown here) we find qualitatively similar results for the energy density fluctuations $\delta\ed$.}
\label{Distributions}\end{figure*}

\subsection{Numerical Results for Distributions}

We choose initial conditions of the Bunch-Davies vacuum for modes that are deep inside the apparent horizon during inflation, and evolve numerically.
The late time power spectrum for dimension 4 potentials is given in Fig.~\ref{Power} for $\Pv=20$. By taking $\lambda>0$ we see the difference between the behavior of the non-resonant adiabatic mode (upper panel) and the resonant isocurvature modes (lower panel). While for $\lambda<0$ (not shown here) we find the opposite behavior. Since the individual power spectra for $\delta\phi$ oscillate, we have time averaged over a cycle to give the late time average value. We also see qualitatively similar behavior for the corresponding power spectra for the densities $\delta\ed$ and $\delta\nd$.

We can use the above probability distributions to draw sample distributions for the densities. To do so we discretize on a cubic lattice. We call the box size $\lbx^3$ and the lattice spacing $\lat$. The discrete set of allowed wave vectors are
\beq
{\bf k} = {2\pi\over\lbx}\left(m_x,m_y,m_z\right)
\eeq
where $m_x,m_y,m_z$ are integers. The maximum value of the wave vector components is $\pi/\lat$, so the maximum value of the integers is $\lbx/(2\lat)$.

The field in position space is a stochastic variable and can be written as
\beq
\Var({\bf x}) = {1\over \lbx^3} \sum_{\bf k} \Var_k\, e^{i{\bf k}\cdot{\bf x}}
\eeq
where the set of ${\bf k}$ is given above. The Fourier coefficients $\Var_k$ are each drawn from the Gaussian distribution
\beq
\mathcal{P}[\Var_k,t] = {1\over\sqrt{2\pi\sigma_k^2}}\exp\left[-{|\Var_k|^2\over 2\,\sigma_k^2(t)}\right]
\eeq
where the variance is
\beq
\sigma_k^2(t) = \lbx^3 P_\Var(k,t)
\eeq
and the reality condition requires $\Var_k^*=\Var_{-k}$.

At very early times, the physical wave numbers of interest are given by the Bunch-Davies vacuum and essentially describe a free field in Minkowski space. We draw on this distribution for the number density $\delta\nd$. The result appears in the left panel of Fig.~\ref{Distributions}. This shows vacuum fluctuations associated with virtual particles/antiparticles. The Minkowski fluctuations are UV sensitive, as the field fluctuations have power that goes as $P\propto 1/k$. This leads to a formally infinite variance for the field in position space $\langle\Var^2\rangle$. We have introduced a UV cutoff of $k_{UV}=3\,a_0\,m$ in this figure.

At late times, a range of modes grow exponentially. This is predominantly for relatively low $k$ modes, as described earlier. For $\lambda>0$ the corresponding power spectra is given in the lower panel of Fig.~\ref{Power}, which shows that there is large power at finite wave numbers with some characteristic scale around $k\sim 0.5\,a_0\,m$. A realization of the number density $\delta\nd$ is given in the right hand panel of Fig.~\ref{Distributions}. 
This shows that for a complex field the inflaton fragments into separate regions of $\phi$-particles and anti-$\phi$-particles. We have plotted $\delta\nd$ in units of $\sqrt{\lambda}\,\nd_0$, where $\nd_0$ is the background density of particles (same as antiparticles) defined as $\nd_0\equiv\ed_0/m$. For $\Pv=20$ we see that the ratio approaches values of several thousand. This is acceptable since the coupling $\lambda$ can be very small, so this can still be in the linear regime where the perturbation $\delta\nd$ is less than the background $\nd_0$. For higher values of $\Pv$ the growth is exponentially larger as discussed in Section \ref{Hubble}. So this can easily lead to large fragmentation of the inflaton, which can lead to large nonlinearities beyond the linear regime.

We also find that qualitatively similar behavior happens for the energy density perturbations $\delta\ed$ when $\lambda<0$.

\section{Application to Baryogenesis} \label{Baryogenesis}

In the previous sections we studied the behavior of the inflaton fields after inflation in a class of models organized by an internal rotational symmetry. 
The internal symmetry means that the field carries a conserved particle number. It is of interest to examine whether this conserved particle number may be associated with the late time conserved baryon number. 

However, as mentioned earlier, the slow-roll inflationary phase will cause the inflaton to evolve radially in field space, and the net particle number associated with this $\Delta\Num=\Num_\phi-\Num_{\bar{\phi}}$ (number of particles minus antiparticles) vanishes. So in order to produce a non-zero net particle number, we need to introduce a breaking of the symmetry. In Refs.~\cite{Hertzberg:2013jba,Hertzberg:2013mba} some of us developed a method to achieve this, as an inflationary version of the classic Affleck-Dine mechanism for baryogenesis. This idea is particularly appealing as it more easily satisfies constraints on cosmological isocurvature fluctuations, that can otherwise be problematic for low scale Affleck Dine models. Some interesting follow up works includes \cite{Lozanov:2014zfa,Takeda:2014eoa,Li:2014jqa}.

In our previous work \cite{Hertzberg:2013jba,Hertzberg:2013mba}, only the homogeneous $\phi_0(t)$ was considered; here we would like to include corrections from the inhomogeneous $\delta\phi({\bf x},t)$ that arises from self-resonance.

\subsection{Inflationary Baryogenesis Models}

Lets focus on the case of a complex inflaton field $\phi$. For canonical kinetic energy and standard gravity, its dynamics are governed by the choice of potential $V(\phi)$. Earlier in this paper, we imposed a global $U(1)$ symmetry on this potential so that it only depends on the magnitude of $\phi$; here we relax this. We decompose the potential in terms of a symmetric piece $V_s$ that respects the $U(1)$ symmetry and an asymmetric piece  $V_b$ that breaks the symmetry
\beq
V(\phi,\phi^*) = V_s(|\phi|)+V_b(\phi,\phi^*)
\eeq
where we have made it explicit that the potential now is a function of 2 variables. We assume that the symmetry is weakly broken. This means that the symmetric piece $V_s$ is the dominant piece of the potential, both during and after inflation, and the asymmetric piece $V_b$ is subdominant. In order to recover the symmetry at late times, as the field redshifts, we assume the symmetric piece includes a (positive) mass term 
\beq
V_s(|\phi|)=m^2|\phi|^2+\ldots
\eeq
where the dots indicate higher order operators, such as $\lambda|\phi|^4$, that can lead to self-resonance; these higher order terms are even allowed to dominate at large field values relevant for inflation. This symmetric potential $V_s$ plays the same role as the symmetric potential $V$ we studied in the earlier sections. At large field values, the higher order terms may organize the potential into one with negative pressure or positive pressure, as we saw earlier. This then determines which mode is resonant at long wavelengths.

Lets assume that the breaking term is dominated by a single operator. We take this to be a power law of the form
\beq
V_b(\phi,\phi^*) = \bre\,(\phi^\pb+\phi^{*\pb})
\eeq
where $\pb\geq3$ is the operator dimension of the $U(1)$ breaking. Since $\pb\ge3$, and the symmetric piece includes the quadratic mass term, the symmetry is indeed restored at late times as the field redshifts to small values. 

A couple of possible justifications of this Lagrangian are as follows: (a) imposing a discrete $\mathbb{Z}_n$ symmetry, (b) promoting $\phi$ to carry color charge; this allows for a color singlet operator that breaks the global $U(1)$ as $\sim \epsilon_{ijk}\phi^i\phi^j\phi^k$ ($\pb=3$) for multiple generations. Each of these justifications has its own advantages and disadvantages. For (a) it nicely organizes the action into the desired form, but leaves open the question as to the origin of this discrete symmetry. In (b) it naturally leads to the $\pb=3$ breaking term, but it is non-trivial to give the inflaton charge since that will tend to renormalize the self couplings of the inflaton. Anyhow, a full analysis of the embedding into microscopic physics is not the focus of the present paper.

Let us turn our attention to the time evolution. The full nonlinear equation of motion (including Hubble expansion) is
\beq
\ddot\phi+3 H\dot\phi-{\nabla^2\phi\over a^2}+{V_s'(\phim)\over\phim}\phi + \pb\,\bre\,\phi^{*\pb-1} = 0
\eeq
(where $\phim=\sqrt{2}\,|\phi|$).
By tracking the evolution of the complex inflaton, we see that the final term causes an alteration to the purely radial motion.
Then, as demonstrated in Refs.~\cite{Hertzberg:2013jba,Hertzberg:2013mba}, this leads to a non-zero particle number, which can later decay to quarks providing a baryon asymmetry. The details of this final decay are model dependent. 

Since (i) the $U(1)$ symmetry associated with baryon number is explicitly broken, (ii) the C and CP symmetries are spontaneously broken by the inflaton's VEV, and (iii) the decay into quarks is out of equilibrium; the Sakharov conditions for baryogenesis are satisfied.

\subsection{Weakly Broken Symmetry Approximation}

At late times when the symmetry is restored, there is a conserved net particle number given by
\beq
\Delta\Num = i\int d^3x\,a^3(t)\!\left[\dot\phi\,\phi^* - \dot\phi^*\,\phi\right]
\eeq
In principle, we could solve the full nonlinear equations of motion for some set of initial conditions, and then integrate over space to obtain $\Delta\Num$. However, we would like to use our previous analysis involving Floquet theory to obtain an approximation to this.

First, let us take a time derivative of this quantity. We note that it will not be conserved due to the presence of the breaking term. It is straightforward to use the equation of motion for $\phi$ to simplify $\dot{\Delta\Num}$. By integrating the result, we obtain
\beq
\Delta N(t_f) = i\,\bre\,\pb\int^{t_f}_{t_i} \!dt\,d^3x\,a^2(t)\left[\phi^\pb({\bf x},t)-\phi^{*\pb}({\bf x},t)\right]\,\,\,\,
\label{DNint}\eeq
Note that it is proportional to the strength of the breaking $\bre$, as it should be. We integrate over time from some initial early time $t_i$ (say the start of inflation, where the number is negligibly small since the comoving volume is so small) to some late final time $t_f$. In fact the answer will asymptote to a constant as $t_f\to\infty$ as the particle number becomes conserved at late times.

Since the expression for $\Delta\Num$ in eq.~(\ref{DNint}) is proportional to $\bre$, then for a sufficiently weak breaking of the symmetry (small $\bre$) we can evaluate the quantity inside the integral to zeroth order in $\bre$. That is to say, we can use the symmetric theory to determine $\phi$ as an input into the integral. This means that {\em the earlier results in this paper on self-resonance in symmetric theories can be utilized here to tell us about the net number of particles produced in asymmetric theories in the weakly broken regime.}

As usual we decompose the field into a background piece and a perturbation. Since we can treat the field as arising from the symmetric theory, we can take the background to undergo radial motion as usual. Lets call the fixed angle in the complex plane of the radial oscillations $\theta_i$. It is then useful to decompose the field into background and perturbations $\phi=\phi_0+\delta\phi$ as follows
\bea
\phi_0(t) &=& {e^{i\theta_i}\over\sqrt{2}} \phim_0(t) \\
\delta\phi({\bf x},t) &=& {e^{i\theta_i}\over\sqrt{2}} \left(\delta\phi_\parallel({\bf x},t)+i\,\delta\phi_\perp({\bf x},t)\right)
\eea
We substitute this into eq.~(\ref{DNint}) and expand to leading non-zero order in $\delta\phi$. 
The result for the net number of particles can be decomposed as
\beq
\Delta\Num = \Delta\Num_0 + \Delta\Num_\delta
\eeq
where $\Delta\Num_0$ is the background contribution and $\Delta\Num_\delta$ is the correction from perturbations. 
The background piece is \cite{Hertzberg:2013jba,Hertzberg:2013mba}
\beq
\Delta\Num_0(t_f) =  -\bre{V_{com}\,\pb\over 2^{{\pb\over2}-1}}\sin(\pb\,\theta_i)\int_{t_i}^{t_f}\!dt\, a^3(t)\,\phim_0^{\pb}(t)
\eeq
where $V_{com}$ is a comoving volume.

\begin{figure}[t]
  \includegraphics[width=\cw]{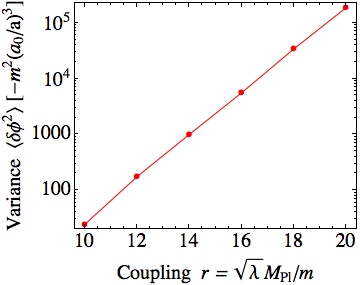}
\caption{The late time difference in variances $\langle\delta\phi^2\rangle\equiv\langle\delta\phi_\parallel^2\rangle-\langle\delta\phi_\perp^2\rangle$ from the exponential growth of IR modes (the UV divergence cancels between the 2 terms) as a function of the dimensionless coupling $\Pv$ for dimension 4 potentials, with $m^2>0$ and $\lambda>0$. We have plotted the variance in units of $m^2(a_0/a)^3$ and absorbed an overall negative sign (since the orthogonal modes are resonant, the difference is negative).}
\label{Variance}\end{figure}

At linear order in the perturbations $\sim\delta\phi$, the contribution to $\Delta\Num$ vanishes. This is because
\beq
\int d^3x\,\delta\phi({\bf x},t) = 0
\eeq
as the zero mode is entirely captured by $\phi_0(t)$, by definition. 
This means that the leading non-zero contribution to $\Delta\Num$ is quadratic in the perturbations $\sim\delta\phi^2$.  It is useful to take the quantum expectation value of this result, leading to variances of the fluctuations.
We find the result
\bea
\Delta\Num_\delta(t_f) &=& -\bre{V_{com}\,\pb\,C(\pb,2)\over 2^{{\pb\over2}-1}}\sin(\pb\,\theta_i)\nonumber\\
&&\times\int_{t_i}^{t_f}\!dt\, a^3(t)\,\phim_0^{\pb-2}(t) \langle\delta\phi^2(t)\rangle
\label{DNMaster}\eea
The expression $\langle\delta\phi^2\rangle$ is shorthand for the difference in the variances $\langle\delta\phi^2\rangle\equiv\langle\delta\phi_\parallel^2\rangle-\langle\delta\phi_\perp^2\rangle$.
The variances may be expressed in terms of integrals over the power spectra as follows
\beq
\langle\delta\phi^2(t)\rangle = \int\!\!{d^3 k\over(2\pi)^3}\!\left[P_{\delta\phi_\parallel}(k,t)-P_{\delta\phi_\perp}\!(k,t)\right]
\label{VarInt}\eeq
where $P_{\delta\phi_\parallel},\,P_{\delta\phi_\perp}$ are the power spectra from Section \ref{PositionSpace} for the symmetric theory. 
A plot of this finite difference in variance is given in Fig.~\ref{Variance}.

Due to statistical isotropy, the 3-dimensional integral over wave-vectors, simplifies to a 1-dimensional integral
\beq
\int\!\!{d^3 k\over(2\pi)^3}\to \int\! {dk\,k^2\over 2\pi^2}
\eeq
Now recall that the dominant exponential instability, if present, is for relatively low wave numbers. On the other hand, the asymptotically high wave numbers are deep in Minkowski space corresponding to mode functions that evolve circularly in the complex plane. So the power spectra at high $k$ are approximated by their Minkowski space values
\beq
P_{\delta\phi_\parallel}(k,t)\approx P_{\delta\phi_\perp}(k,t)\approx {1\over 2\,k\,a^2}
\eeq
Hence each of the individual terms inside the $k$-integral in eq.~(\ref{DNMaster}) would give rise to a quadratic UV divergence. However, the {\em difference is finite}. A non-zero difference primarily arises from the finite $k$ regions in the Floquet chart that carry an instability.

\subsection{Numerical Results for Baryon Asymmetry}\label{BaryonNumerical}

We would like to report on results for the particle/antiparticle asymmetry. A useful measure of asymmetry comes by defining the asymmetry parameter as the difference between the particle and antiparticles numbers $\Delta\Num=\Num_\phi-\Num_{\bar{\phi}}$ divided by their sum
\beq
\As \equiv {\Num_\phi-\Num_{\bar{\phi}}\over \Num_\phi+\Num_{\bar{\phi}}} =  {\nd_\phi-\nd_{\bar{\phi}}\over \nd_\phi+\nd_{\bar{\phi}}}
\eeq
where in the latter expression we have recast this in terms of densities. Now the difference is well defined as it is associated with a conserved quantity in the weakly broken limit. However the sum is in general not well defined in a relativistic theory. However at late times, we enter the nonrelativistic regime, where it is given through the energy density as $\nd_\phi+\nd_{\bar{\phi}}=\ed/m$.

Let us begin with the homogeneous approximation. Using the above expressions, we can write this as 
\beq
A_0 = - c_n\,\bre\,{m^{\pb-4}\over \lambda^{{\pb\over2}-1}}\sin(n\,\theta_i)
\label{AsymmP}\eeq
where
\beq
c_n = {\pb\over 2^{\pb\over2}-1}{\int \! d \td\,a(\td)^3\phim_d(\td)^\pb\over a^3\,\ed_d}
\eeq
is a type of ``asymmetry coefficient". Here $\td\equiv m\,t$, $\phim_d\equiv\sqrt{\lambda}\,\phim_0/m$, and $\ed_d\equiv \lambda\,\ed/m^4$ is a dimensionless time, field, and energy density, respectively. The end point of integration and the denominator is to be evaluated at late times.
Note that the ratio of couplings in eq.~(\ref{AsymmP}) should be small. For $\pb=3$, the ratio is $\bre/(m\sqrt{\lambda})$; this should be small so that the breaking term is always subdominant to the quadratic or quartic terms. For $\pb=4$, the ratio is $\bre/\lambda$; this should be small so that the asymmetric quartic term is subdominant to the symmetric quartic term.

\begin{figure}[t]
  \includegraphics[width=\cw]{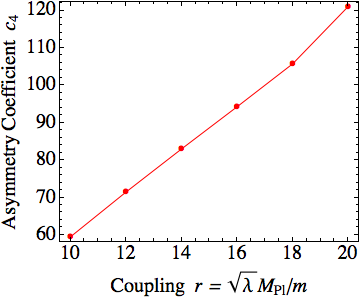}
\caption{The coefficient $c_4$ that controls the homogeneous asymmetry as a function of the dimensionless coupling $\Pv$ for dimension 4 potentials, with $m^2>0$, $\lambda>0$, and $\pb=4$.}
\label{AsymmetryParameter}\end{figure}

We have numerically computed the above integral to determine the asymmetry coefficient $c_n$. 
As an example, we give the result in Fig.~\ref{AsymmetryParameter} for standard dimension 4 inflationary potentials with $\pb=4$. This shows that the asymmetry grows relatively mildly as we increase $\Pv$.

\begin{figure}[t]
  \includegraphics[width=\cw]{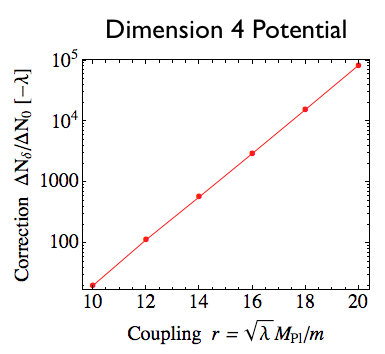}\\\vspace{0.5cm}
    \includegraphics[width=\cw]{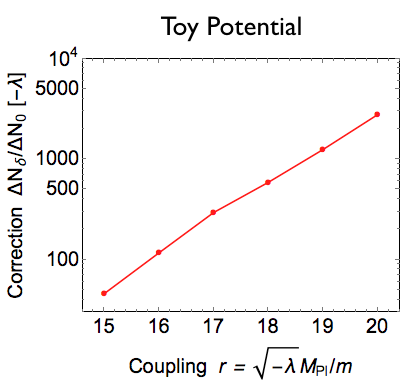}
\caption{The relative correction to the baryon asymmetry $\Delta N_\delta/\Delta N_0$ (quantum fluctuations relative to classical background) as a function of the dimensionless coupling $\Pv$ for two different kinds of inflationary potentials with $\pb=4$. Upper panel is for the dimension 4 potential (with $m^2>0$ and $\lambda>0$). Lower panel is for the toy potential of eq.~(\ref{ModPot}) (an $m^2>0$ and $\lambda<0$ model).
We have plotted the relative correction in units of $\lambda$ and absorbed an overall minus sign.}
\label{Correction}\end{figure}

The correction from parametric resonance of perturbations arises from computing the integral over time and wave numbers in eqs.~(\ref{DNMaster},\,\ref{VarInt}). We have carried out these integrals with the result given in upper panel of Fig.~\ref{Correction} for standard dimension 4 inflationary potentials. We have plotted the asymmetry correction $\Delta\Num_\delta$ in units of the homogeneous asymmetry value $\Delta\Num_0$ and rescaled by $\lambda$ and absorbed an overall minus sign. This shows that for these parameters, the asymmetry is reduced (due to the overall minus sign) relative to the homogeneous approximation. Also, we see there is exponential sensitivity to the parameter $\Pv$. So for large $\Pv$ a full nonlinear treatment would be useful to take into account effects of back reaction. Nevertheless the numerical results here give a sense of some of the overall qualitative behavior.

For negative quartic coupling we need to regulate the potential at large field values. As a concrete example to illustrate the possibilities, we consider the following toy potential
\bea 
V_s(|\phi|) = {{1\over2}m^2|\phi|^2\over \sqrt{1+|\phi|^2/\F^2}}
\label{ModPot}\eea
and we take $m^2>0$.
When Taylor expanded around $\phi=0$, this gives a positive mass and negative quartic term with $\F=m/\sqrt{|\lambda|}$. The potential grows more slowly than a quadratic, namely as $V_s\sim |\phi|$ at large field values, which supports a phase of slow-roll inflation. Another class of toy potentials was mentioned in Part 1 Section V B 5 \cite{Part1} for $0<\ex<1$; those potentials are motivated by string axions \cite{McAllister:2008hb,ModMcAllister}, which do not carry baryon number, so we do not study them further here. With potential (\ref{ModPot}) we numerically solve for the baryon asymmetry, with the result given in the lower panel of Fig.~\ref{Correction}. In this case the correction is enhanced relative to the background value.

\section{Conclusions} \label{Conclusions}

In this paper we have further developed a theory of self-resonance after inflation from Part 1 \cite{Part1}. 
We have explained the deep reason for the self-resonance behavior in terms of the underlying description of the quantum mechanics of many particles.

In the nonrelativistic regime re-organized the theory into contact interactions between particles and antiparticles, with coupling strength given by $\lambda$. For $\lambda>0$ the particles (and antiparticles) exhibit repulsion, so the homogeneous configuration established by inflation is stable against adiabatic perturbations. While for $\lambda<0$ there is mutual attraction leading to breakup of the homogeneous configuration and instability. On the other hand, the isocurvature modes have very different behavior from the adiabatic modes. In particular, for $\lambda>0$ the isocurvature mode leads to an instability, despite the repulsion among the particles (and antiparticles). The reason is that Bose-Einstein statistics favor the particles to clump with particles and the antiparticles to clump with antiparticles.

We also developed a small amplitude analysis, which captured not only the long wavelength behavior, but also the higher band structure, and we explained this in terms of Feynman diagrams of annihilation and decay, where appropriate.

We then performed the quantization of our perturbations. As an example we computed the distribution of the number density of particles minus the density of antiparticles for $\lambda>0$. For strong resonance, we showed that the inflaton fragments into separate regions of particles and antiparticles.

Finally, we applied the quantization of the inflaton fields to the case of particle-antiparticle asymmetry, which is relevant to some models of baryogenesis \cite{Hertzberg:2013jba,Hertzberg:2013mba}. We showed that the symmetric theory can be used to compute the leading order behavior of the asymmetric theory in the limit of weak symmetry breaking. We computed the corrections to the homogeneous theory from the inhomogeneous theory due to self-resonance. The result involves an integral over the difference in the power between the adiabatic modes and the isocurvature modes. This difference is finite and is dominated by the resonant modes at relatively long wavelengths.

Altogether, along with Part 1 \cite{Part1}, our work gives a detailed theory of self-resonance after inflation in single and multi-field models. In Part 1 \cite{Part1} we understood the long wavelength behavior using the Goldstone theorem. Here we have provided the deep underlying physical understanding of the entire resonance structure, with the main structure determined by the attraction/repulsion of particles. 

An interesting direction for future work is to incorporate corrections from possibly large fluctuations of the metric, and explore gravitational waves production. Another direction is to extend this theory by including couplings to other fields beyond the inflaton, such as Standard Model fields.

\bigskip

\begin{center}
{\bf Acknowledgments}
\end{center}
We thank Jacovie Rodriguez for help in the early stage of this project. We also thank Alan Guth, David Kaiser, Mustafa Amin, and Kaloian Lozanov for useful discussions.
We would like to acknowledge support by the Center for Theoretical Physics and the Undergraduate Research Opportunities Program at MIT. 
This work is supported by the U.S. Department of Energy under cooperative research agreement Contract Number DE-FG02-05ER41360. 
JK is supported by an NSERC PDF fellowship.


\end{document}